\documentclass{emulateapj}
\usepackage{natbib}

\newcommand{\Msun}{$M_{\sun}$}

\begin{document}
\title{High Resolution Molecular Gas Maps of M33}
\shortauthors{Rosolowsky et al.}
\shorttitle{GMCs and CO in M33}

\author{Erik Rosolowsky}
\affil{Harvard-Smithsonian Center for Astrophysics, 60 Garden Street,
MS-66, Cambridge, MA 02138} 
\email{erosolow@cfa.harvard.edu}
\author{Eric Keto}
\affil{Harvard-Smithsonian Center for Astrophysics, 60 Garden Street,
MS-78, Cambridge, MA 02138} 
\email{eketo@cfa.harvard.edu}
\author{Satoki Matsushita}
\affil{Academia Sinica Institute of Astronomy and Astrophysics, 
P.O. Box 23-141, Taipei 106, Taiwan}
\email{satoki@asiaa.sinica.edu.tw}
\author{S.~P.~Willner}
\affil{Harvard-Smithsonian Center for Astrophysics, 60 Garden Street,
MS-65, Cambridge, MA 02138} 
\email{swillner@cfa.harvard.edu}

\begin{abstract}
New observations of CO ($J=1\to 0$) line emission from M33, using the
25 element BEARS focal plane array at the Nobeyama Radio Observatory
45-m telescope, in conjunction with existing maps from the BIMA
interferometer and the FCRAO 14-m telescope, give the highest
resolution ($13''$) and most sensitive ($\sigma_{rms}\sim 60$~ mK)
maps to date of the distribution of molecular gas in the central 5.5
kpc of the galaxy.  A new catalog of giant molecular clouds (GMCs) has
a completeness limit of $1.3\times10^5$~\Msun.  The fraction of
molecular gas found in GMCs is a strong function of radius in the
galaxy, declining from 60\% in the center to 20\% at galactocentric
radius $R_{gal}\approx4$~kpc.  Beyond that radius, GMCs are nearly
absent, although molecular gas exists.  Most (90\%) of the emission
from low mass clouds is found within 100 pc projected separation of a
GMC.   In an annulus $2.1<R_{gal}<4.1$~ kpc, GMC masses follow a
power law distribution with index $-2.1$.  Inside that radius, the
mass distribution is truncated, and clouds more massive than
$8\times10^5$~\Msun\ are absent.  The cloud mass distribution shows no
significant difference in the grand design spiral arms versus the
interarm region.  The CO surface brightness ratio for the arm to
interarm regions is 1.5, typical of other flocculent galaxies.
\end{abstract}
\section{Introduction}

The nearby galaxy M33 is an ideal site to study molecular gas and star
formation in the larger context of a disk galaxy.  Most of our
knowledge of star formation and molecular clouds is informed by nearby
star forming regions such as Taurus and Orion, but studies of star
formation across the Milky Way as a whole are confused by our
perspective from within the Galactic disk.  In contrast, the M33 
disk face is visible \citep[$i\approx 52^{\circ}$,][]{cs00}, so
it is easier to study molecular gas in relationship to other
components of the galaxy.  In addition, M33 is near enough
\citep[$D=840$~kpc,][]{Freedman2001} that individual molecular clouds
can be resolved with a large single-dish telescope or millimeter
interferometer.

Because of the perspective that observations of M33 offer, the galaxy
has been the target of many large-scale studies of molecular gas.
\citet{ws89} observed the inner region of the galaxy in  
CO~($J=1\to 0$) with the NRAO 12-m and followed up with high
resolution observations using the Owens Valley interferometer
\citep{ws90}.  The high resolution observations were used to study the
properties of individual molecular clouds and compare to clouds
in the Milky Way.  A further study \citep{wwt97} took
advantage of the disk visibility to study variations in the
properties of molecular gas across a range of galactocentric radii.
Following improvements in millimeter-wave instrumentation, large-scale
surveys of the galaxy became possible, such as that of
\citet[][hereafter EPRB]{eprb03} who used the BIMA millimeter interferometer
to identify all GMCs across the star forming disk.
\citet{rpeb03} studied a limited area of the disk, comprising about
1/3 of the clouds in the BIMA survey, at higher resolution.  The study
determined individual cloud properties including virial masses,
confirmed the work of \citet{ws90}, and constrained GMC formation
mechanisms by comparing the predicted and observed values for GMC
angular momentum.  Finally, \citet[][hereafter HCSY]{m33-fcrao} used
the FCRAO 14-m to complete a single-dish survey of the entire galaxy.
Comparison with {\it IRAS} data determined the local star formation
rate as a function of the surface density of molecular gas.

The existing observations, while providing a wealth of data, also
raise additional questions that merit another look at the molecular
gas in M33.  In particular, comparison of the interferometer-only
study of EPRB with the single-dish survey of HCSY shows that the
interferometer recovered only 20\% of the CO flux in the galaxy!
Because the typical size scales of molecular clouds in M33 should
roughly match the synthesized beam of the interferometer, such a
small fraction of flux recovery is unexpected. What is the nature of
the remaining molecular gas?  One clue comes from the steep mass
distribution of GMCs in M33. Unlike the Milky Way, EPRB found that
the mass distribution was ``bottom heavy'' i.e. most of the molecular
mass was found in clouds near their sensitivity limit.  Thus, it
appears that most of the missing flux is comprised of low-mass
molecular clouds that the interferometer could not detect. This
prompts several questions about the nature of such clouds: Where are
these low mass molecular clouds located relative to the GMCs?  Does
the fraction of material found in GMCs vary over the galaxy?  These
questions are not easily answered by the existing observations,
because the diffuse, extended emission of the lower-mass clouds is
not easily detected by a radio interferometer, and the existing
single dish observations do not have the angular resolution to define
individual clouds.

To study the distribution of low mass clouds in more detail, we
mapped the inner 2~kpc of M33 using the BEARS receiver on the 45-m
telescope at the Nobeyama Radio Observatory (NRO).  The NRO 45-m data give
better sensitivity than previous observations and allow detection of
a greater fraction of the CO in the central region.  We present two
new maps: a BIMA+FCRAO map of the entire galaxy and a BIMA+FCRAO+NRO
map of the inner 2~kpc.  The maps are complementary --- the first
surveys the entire galaxy at $13''$ (50~pc) resolution, sufficient to
identify individual GMCs, while the second has 50\% better point
source sensitivity and a factor of 3.5 better column density
sensitivity, though having poorer resolution ($20''= 75$~pc) and
limited coverage.
Section~\ref{obs} of this paper describes
the new observations at NRO, while \S\ref{redux} explains
the data processing and  production of the maps.
Section~\ref{gmcs-again}
presents the new GMC catalog.  The new catalog demonstrates
significant variations in the mass distribution across the face of
the galaxy, as described in \S\ref{massspec}.  Section~\ref{summ}
summarizes the results.


\section{Nobeyama Radio Observatory Observations}
\label{obs}

The BEARS array on the 45~m telescope at Nobeyama Radio Observatory
consists of 25 SIS mixer receivers arranged in a $5\times 5$ array
with 41\farcs1 between the elements \citep{bears-sunada}.  With the
$14''$ beam size at 115~GHz, the elements are separated by 3
resolution elements on the sky.  The basic unit of the M33
observations consisted of nine separate pointings arranged in a
$3\times 3$ grid, each offset from the previous by 14\arcsec.  This
strategy observes a $(3.5')^2$ region at half-Nyquist sampling.  To
acquire fully sampled data, this observation must be repeated three
additional times with a 7\arcsec\ offset in each direction.  Thus, a
fully-sampled map consists of 900 individual pointings separated by
$7''$; a half-sampled map contains 225 pointings separated by $14''$.
The BEARS receiver requires using the 25-element autocorrelator array
as a back end. It has a total bandwidth of 512~MHz split into 1024
channels, resulting in a nominal velocity resolution of
1.3~km~s$^{-1}$ in the $^{12}$CO~($J=1\to 0$) line.

\begin{figure}
\plotone{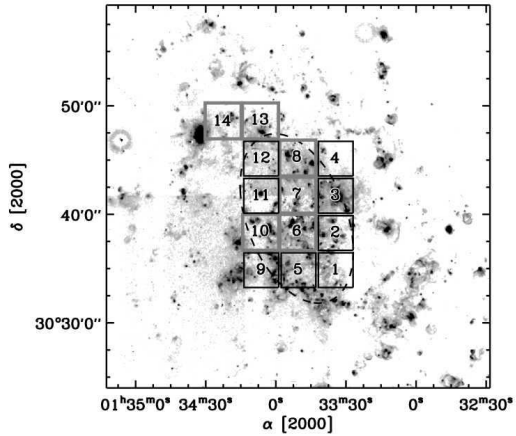}
\caption{\label{obspos} M33 field locations observed with the
NRO 45-m.  The boundaries of the observed fields established by the
3.5$'$ size of the BEARS receiver are indicated on an H$\alpha$ map of
the galaxy \citet{lgs-m31m33}.  Of the 14 fields shown, the six fields with
grey borders were observed with full sampling and the remaining 8
fields were sampled half as densely.  A dashed ellipse is drawn at
$R_{gal}=2$~kpc, illustrating that the survey covers the galaxy out to
this distance.  Properties of the fields are given in Table
\ref{fieldprops}.}
\end{figure}

The observation strategy divided M33 into $(3.5')^2$ regions defined
by the footprint of the receiver array on the sky.  Figure
\ref{obspos} shows the fields observed, and Table~\ref{fieldprops}
gives their positions.  For each field observed with
half-Nyquist sampling, we observed the nine individual position
offsets in 20 second scans, then position switched to an absolute
reference position.  The reference positions were chosen to be
\ion{H}{2} regions at large galactocentric radius ($R_{gal}$).  These
regions may contain CO emission, but they were always chosen so that
the velocity of any emission in the reference beam would be much
different from the line velocity in the observation position.  In good
observing conditions, two scans could be obtained between reference
positions.  In marginal weather, the number of scans had to be
decreased to one, and the reference position was an offset of $30'$ in
azimuth instead of an absolute position.  One pass through the nine
positions of a field took five to six minutes depending on the settle
time for the telescope, and typical observations utilized 12 cycles
obtaining 180 seconds of integration time at each position.  This
completed a half-Nyquist observation of a field in 90 minutes.  Before
and between each field, we updated the pointing solution for the
telescope using the S40 receiver to observe SiO ($J=1\to 0,v=0,1$)
maser emission from IRC+30021.  In low wind conditions, the pointing
solution drifted by $<5''$ over the hour required to observe a single
field.  When the wind velocity was steadily above 5~m~s$^{-1}$, the
pointing drift could exceed $10''$ over an hour.

\begin{deluxetable}{ccccc}
\tablecaption{\label{fieldprops}NRO/BEARS Field Parameters}
\tablewidth{0pt}
\tablehead{
\colhead{Field} & \colhead{Central Position\tablenotemark{a}} &
\colhead{Scans\tablenotemark{b}} & \colhead
{Frac. Retained\tablenotemark{c}} & Sampling}
\startdata
1 & 01:33:34.4  +30:34:51.2 & 108 & 0.94 & Half-Nyq. \\
2 & 01:33:34.4  +30:38:23.6 & 108 & 0.99 & Half-Nyq. \\
3 & 01:33:34.4  +30:41:42.2 & 108 & 0.97 & Half-Nyq. \\
4 & 01:33:34.4  +30:45:07.7 & 108 & 0.84 & Half-Nyq. \\
5 & 01:33:50.9  +30:34:51.2 & 108 & 0.96 & Half-Nyq. \\ 
6 & 01:33:50.9  +30:38:23.6 & 382 & 0.71 & Nyquist \\
7 & 01:33:50.9  +30:41:42.2 & 324 & 0.84 & Nyquist \\
8 & 01:33:50.9  +30:45:14.5 & 323 & 0.87 & Nyquist \\
9 & 01:34:06.8  +30:34:51.2 & 112 & 0.95 & Half-Nyq. \\
10 & 01:34:06.8  +30:38:23.6 & 436 & 0.73 & Nyquist \\
11 & 01:34:06.8  +30:41:42.2 & 108 & 0.96 & Half-Nyq. \\
12 & 01:34:06.8  +30:45:07.7 & 108 & 0.97 & Half-Nyq. \\
13 & 01:34:06.8  +30:48:40.0 & 320 & 0.71 & Nyquist \\
14 & 01:34:22.8  +30:48:40.0 & 163 & 0.46 & Nyquist \\
\enddata
\tablenotetext{a}{J2000}
\tablenotetext{b}{Number of 20 second scans completed by the BEARS array.}
\tablenotetext{c}{Fraction of scans retained after automated scan rejection.}
\end{deluxetable}

We observed from 2005 February 9 to February 17 as long as M33 was at
suitable elevation, excluding time lost to maintenance and bad
weather.  In that time, we completed observations of 14 of the
$(3.5')^2$; six fields are Nyquist sampled and the remaining eight
fields are half-Nyquist sampled (see Figure \ref{obspos} and
Table~\ref{fieldprops} for details).

Since each element of the BEARS receiver consists of a double side
band SIS receiver, we calibrated the array with observations using a
single-sideband receiver to obtain an accurate intensity scale.  At
the latitude of the Nobeyama Radio Observatory, M33 transits above the
elevation limit of the 45-m telescope ($80^{\circ}$) rendering it
unobservable for 90 minutes in the middle of the observations.  During
that time we calibrated the efficiency of each pixel of the BEARS
receiver by comparing observations of NGC 7538
($\alpha_{2000}=23^\mathrm{h} 11^\mathrm{m} 45\fs 5$
,$\delta_{2000}=+61^{\circ} 28' 09''$) made with each element of the
array to those made of the same position with the single-sideband S100
receiver.  The intensity scaling factor for each BEARS pixel relative
to the S100 receiver was the ratio of the integrated intensities in
the respective observations.  The BEARS spectra were multiplied by
these scaling factors averaged over the course of the observing run.
The scaling factors ranged from 1.7 to 3.6 with a median value of 2.4.
Daily variation of the scaling factors was $\lesssim 20\%$ for all
elements; this variation dominates the uncertainty in the overall flux
calibration.  This scaling places BEARS spectra onto the S100
intensity scale, and the spectra must be scaled up by a further factor
of 2.56 ($=1/\eta_{mb}$) to correct for the main beam efficiency.
With all corrections, the noise level of the final map is 85 mK on the
$T_A^*$ scale.

\section{Data Reduction}
\label{redux}
\subsection{Nobeyama Radio Observatory Data}

Inspection revealed several pathologies in the NRO/BEARS data. These
are: (1) pointing errors induced by high wind speeds, (2) baseline
variations, and (3) interloper signals in the base band.  Because of
the data volume --- 62300 raw spectra --- automated processing was
needed to correct these problems.  Scans taken when wind velocities
exceeded 7.5~m~s$^{-1}$ were rejected in their entirety.  Spectral
baselines were established for every spectrum by a fourth-order
b-spline fit with break points established every 100~km~s$^{-1}$,
ignoring regions within 32 channels of the beginning and end of the
spectrum, within 20 channels of a bad correlator section near channel
300, and within 30~km~s$^{-1}$ of the \ion{H}{1} line-of-sight
velocity at observed and reference positions in M33.  These exclusions
prevented the routine from fitting a baseline to actual CO emission in
M33\footnote{The \ion{H}{1} velocities were determined as a function
of position based on the orientation parameters and rotation curve of
\citet{cs00}: inclination 52\degr, major axis position angle
22\degr.}.  The baseline fit accounts for low-order baseline ripples
of arbitrary shape but cannot account for high-order ``standing wave''
patterns in the resulting data caused by interloper signals in the
base band.  Spectra affected by such signals can be readily identified
because the rms residual from the b-spline fit is dramatically larger
than for the remainder of the observations.  The small number of
spectra for which the rms residual is 4 times the median were
rejected.  Table~\ref{fieldprops} reports the fraction of scans that
remain after wind speed and baseline rejection.  The acceptable
spectra were Hanning smoothed and down-sampled to a final channel
width of $2.6$~km~s$^{-1}$ before being gridded into a data cube using
a Gaussian smoothing kernel with a FWHM of $20''$.

\subsection{Merging the Interferometer and Single Dish Maps}


The first step in the analysis was to merge the FCRAO (HCSY) and BIMA
(EPRB) data sets.  As noted in \S1, the interferometer map recovers
only $\sim$~20\% of the flux found in the single-dish map, though the
interferometer map detects nearly all the GMCs in the galaxy.  The
14-m FCRAO dish recovers all flux at the small spatial frequencies not
sampled by the BIMA observations, which have minimum baselines
$\lesssim$~7~m.  Therefore, the two maps can be combined to produce a
single high-resolution, fully-sampled map.  The power in the
overlapping range of spatial frequencies shows that the flux scales of
the single-dish and interferometer maps are consistent within their
uncertainties, and no relative flux scaling is required.

\begin{figure*}
\plotone{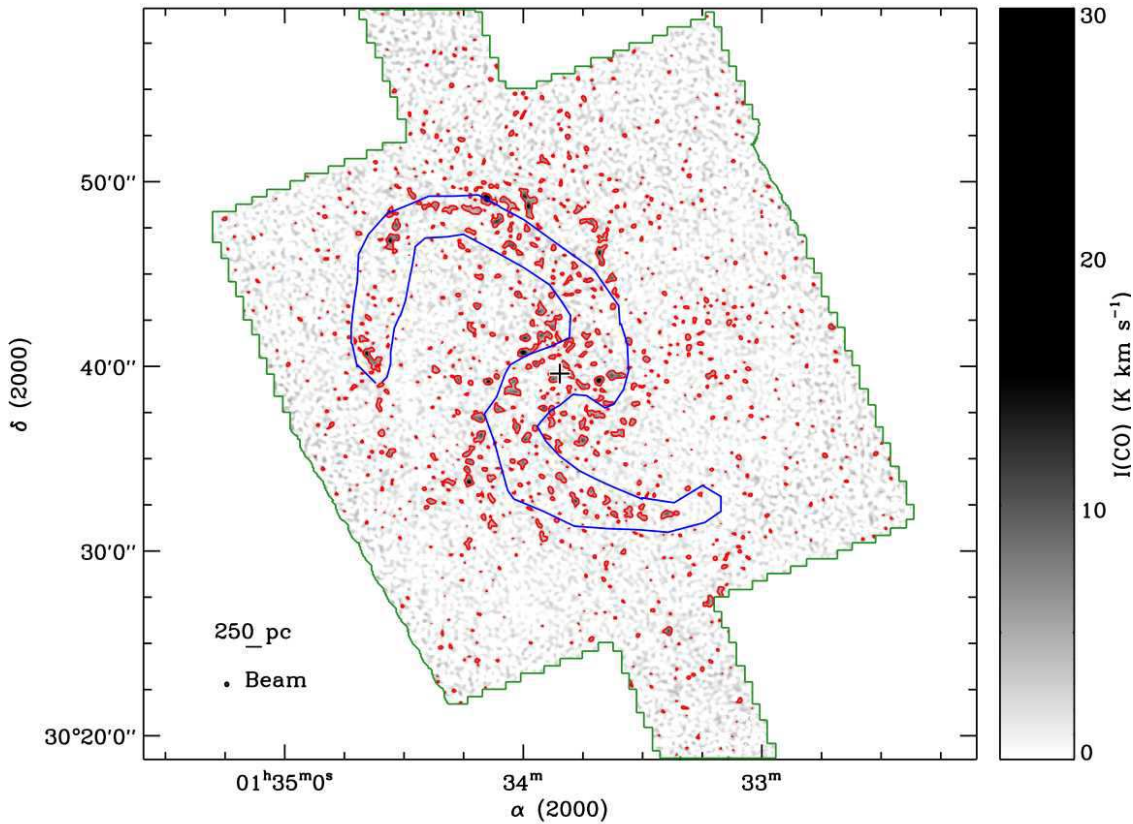}
\caption{\label{comap2} Map of the integrated intensity of CO emission
across M33 from the merged BIMA+FCRAO map.  Intensity at each position
is integrated over a 10 channel (20.3~km~s$^{-1}$) window centered on
line-of-sight projection of the rotation velocity at that position. To
limit the visual confusion, only a single red contour is drawn at a
level of three times the value of the local noise.  Since there are
over $2\times 10^4$ independent positions in the map, some of the
emission above this contour level will be due to statistical
fluctuations alone.  Section \ref{gmcs-again} discusses how spectral
information can be used to distinguish real features from statistical
outliers. The green boundary demarcates the border of the region
observed with both BIMA and FCRAO.  Only in this overlapping region
does the map contain information at all spatial frequencies.  The
center of the galaxy at
$\alpha_{2000}=1^\mathrm{h}~33^\mathrm{m}~50\fs8$,
$\delta_{2000}=+30^{\circ}~39'~36\farcs 7$ is marked with a cross.
The boundaries of the spiral arms adopted in \S\ref{sparm} are shown
in blue.}
\end{figure*}

To assess systematic uncertainties in the resulting map, the data sets
were combined using two techniques: (1) image-domain linear
combination as described by \citet{lincom} and implemented by
\citet{song2} and (2) Fourier-domain combination \citep{immerge}
implemented in MIRIAD \citep{miriad}.  The two methods gave results
indistinguishable within the uncertainties.  Figure \ref{comap2} shows
the result produced by the linear method, which is used throughout
this paper.  This map will be called the ``the merged map.'' Its final
resolution is $13''\times 2.03\mbox{ km s}^{-1}$ (a projected beam
size of 53 pc), and its median noise level is 240~mK. The boundaries
of the FCRAO and BIMA maps do not exactly coincide, but the region of
overlap (shown in Figure \ref{comap2}) covers the majority of the galaxy to
a radius of $R_{gal}\approx 5.5$~kpc.

\subsection{Combining All Data}

\begin{figure*}
\plottwo{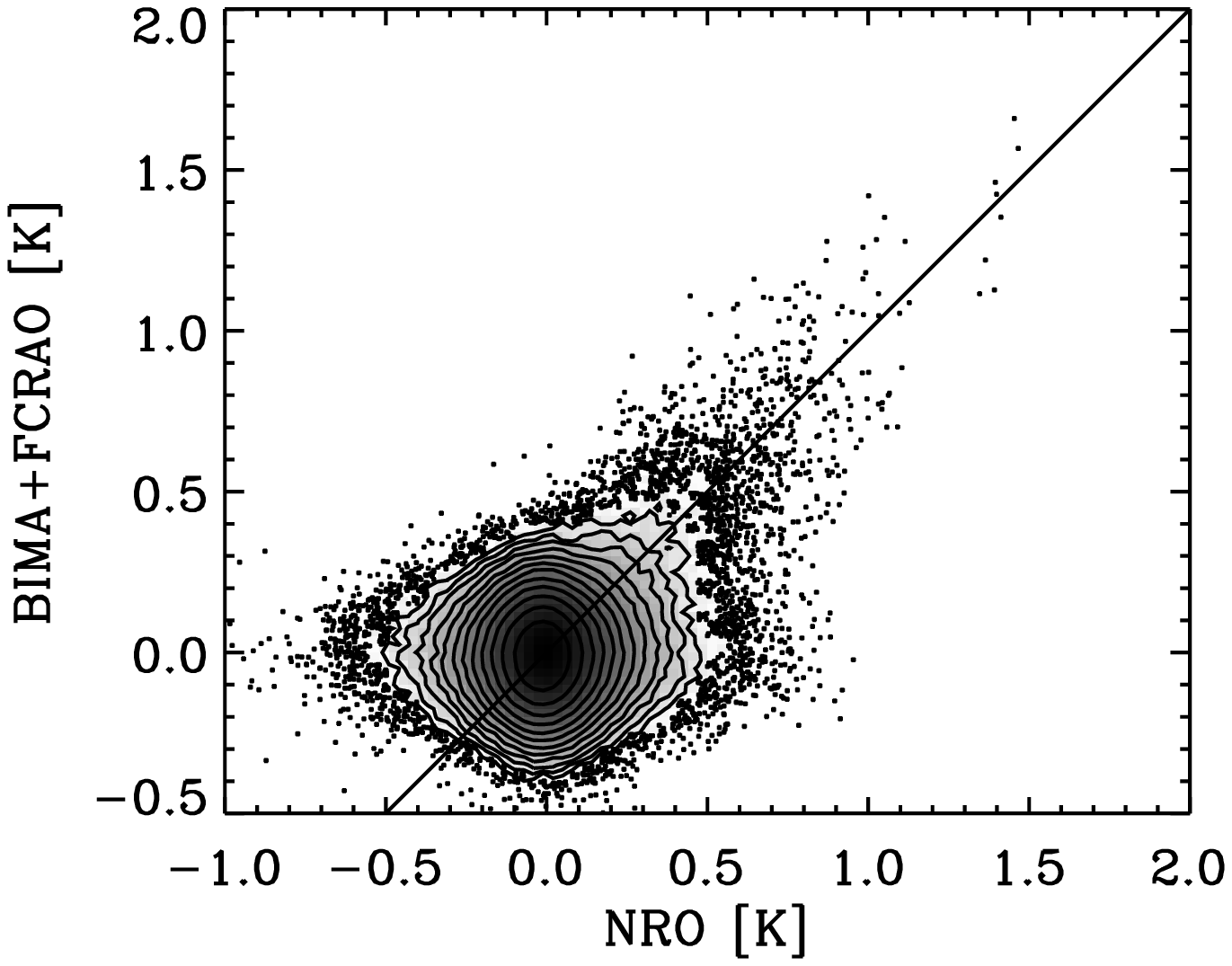}{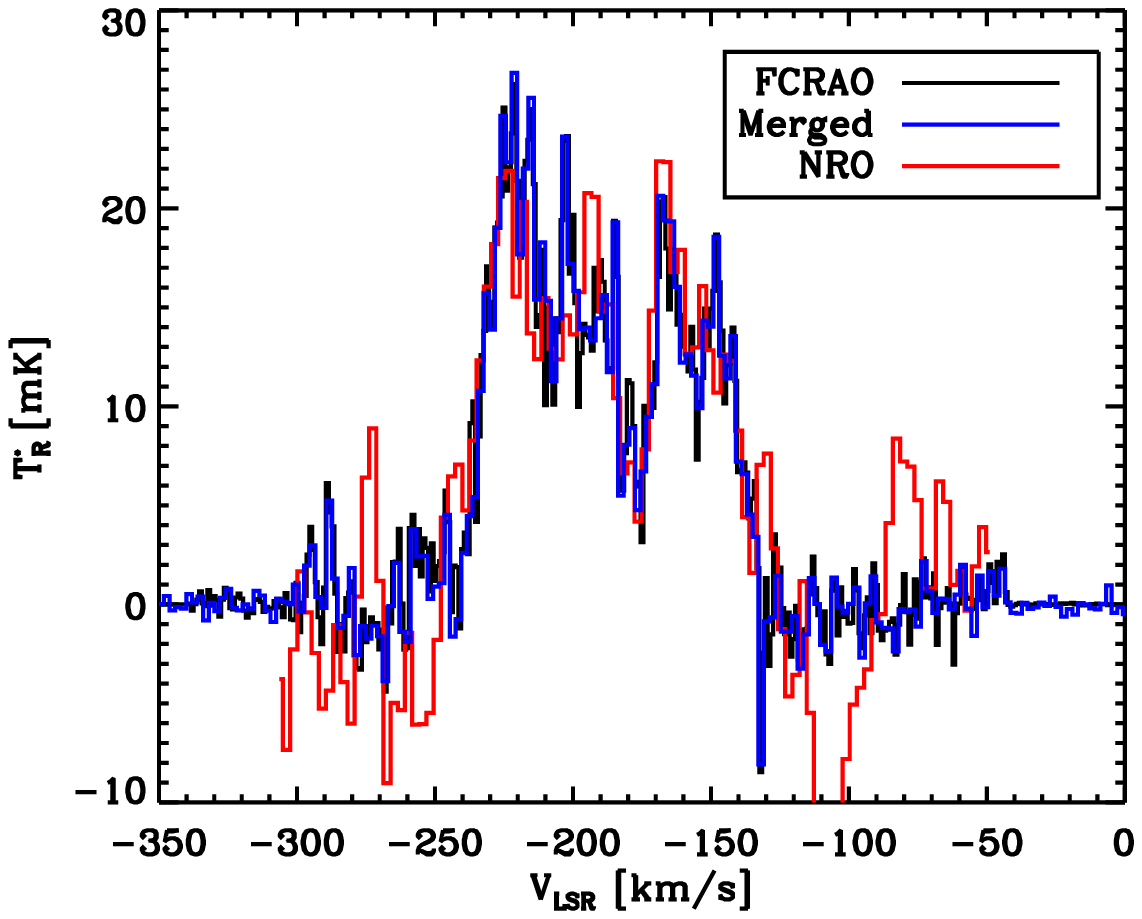}
\caption{\label{datacomp} Comparisons of the merged BIMA+FCRAO map
with the NRO observations. ({\it left}) This panel compares the
individual data values over the shared region between the two maps on
an element-by-element basis showing reasonably good agreement within
the uncertainties of the map.  ({\it right}) This panel compares the
average spectrum that would be derived for each data cube using a
Gaussian beam with FWHM of $7'$ (1.7 kpc) centered on the central
position of the galaxy.  The three spectra are for the merged
BIMA+FCRAO data cube (blue), the FCRAO data set alone (black), and the
NRO data (red). }
\end{figure*}

The NRO 45-m map shows no systematic differences from the merged map.
Figure~\ref{datacomp} compares the two maps in the region of overlap
after convolving the merged map to match the 20$''$ resolution of the
NRO map.  The noise in the NRO map shows greater variation with
position than in the merged map, and therefore many positive and
negative outliers are seen in the pixel comparison plot
(Figure~\ref{datacomp}, left).  Despite that, the concentration of
points near the locus of equality shows agreement in the relative
calibration between the two maps.  The spectral comparison of the
emission from the central $7'$ (1.7 kpc) of the galaxy
(Figure~\ref{datacomp}, right) highlights the agreement in the flux
calibration across the observing band.  The plot also shows that the
combination with the BIMA data does not significantly affect the
amount of flux found in the merged data set.  The agreement between
all three data sets is quite good except near
$V_{LSR}=105$~km~s$^{-1}$ where emission in the reference position of
the NRO 45-m data produces a negative, though irrelevant, feature in
the data.

\begin{figure}
\plotone{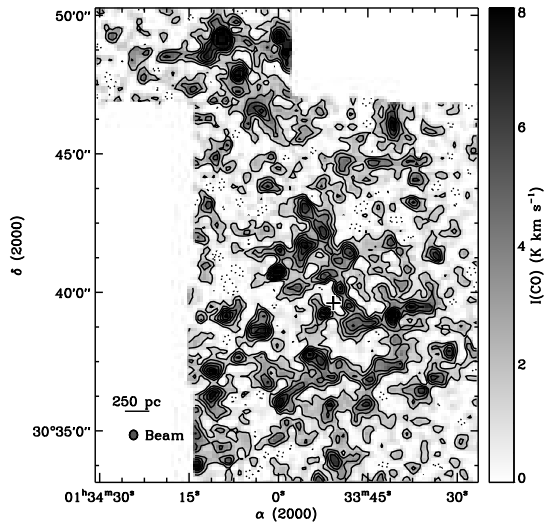}
\caption{\label{comap} Integrated intensity of CO emission in the
central region of M33 from the combined NRO+BIMA+FCRAO data cube.  At
each position, the data have been integrated over a 20 channel
(52~km~s$^{-1}$) wide window centered on the line-of-sight projection
of the rotation velocity at that position as derived from the rotation
curve of \citet{c03}, and the result is shown in negative grey
scale. The noise varies across the map because of variable integration
time and weather, and contours show the statistical significance based
on local estimates of the noise level.  Contour levels are $-4,-2$
(dotted),2,4...14 times the local rms.  The center of the galaxy is
indicated with a cross and is itself devoid of CO.}
\end{figure}

\begin{figure}
\plotone{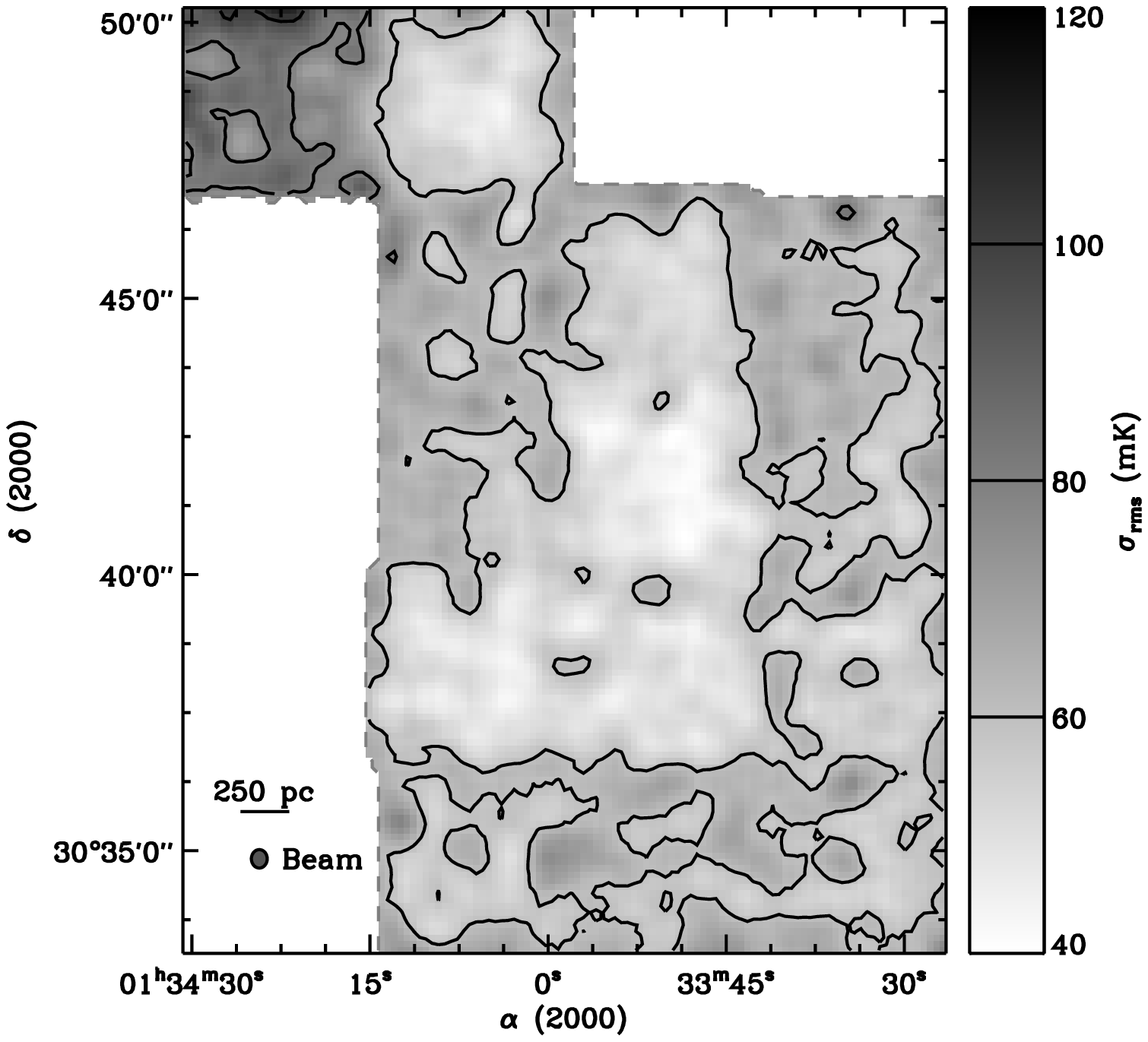}
\caption{\label{noisemap} Map of the 1$\sigma$ rms noise in the combined
NRO+BIMA+FCRAO data cube.  The noise varies because of differences in
integration time across the map and changing atmospheric conditions
during the observing.  Dark areas indicate higher noise as shown by
the scale bar on the right.  Contours are drawn at noise levels of
60, 80, and 100~mK}
\end{figure}

Because the two data sets agree well over their common area, a final
map of the central region of M33 requires simply averaging the two
data cubes.  The average was weighted by the inverse square of the
noise estimate determined for each pixel in the data cube where the
Chauvenet criterion \citep{taylor} was applied iteratively to remove
high-significance outliers (see EPRB for details).  The final data
cube has a typical rms noise temperature of 60~mK and a resolution of
20$''$ $\times$ 2.6~km~s$^{-1}$, corresponding to a projected
resolution of 81 pc.  At this angular resolution, the noise level of
of the BIMA+FCRAO map is 100 mK, so combining the data yields a 40\%
improvement in the sensitivity.  The NRO+FCRAO+BIMA map is shown in
Figure~\ref{comap}; it will be referred to here as ``the combined
map.''  Because of variable integration time and atmospheric
conditions, the noise varies spatially across the map.
Figure~\ref{noisemap} shows the 1$\sigma$ rms noise level as a
function of position for the final data cube.  Table~\ref{cubecomp}
compares the properties of the various maps.

\begin{deluxetable*}{cccc}
\tablecaption{\label{cubecomp} Map Properties}
\tablewidth{0pt}
\tablehead{
\colhead{} & \colhead{FCRAO}& \colhead{BIMA+FCRAO} & \colhead{BIMA+FCRAO+NRO} \\
\colhead{} & \colhead{single dish}& \colhead{``merged''} & \colhead{``combined''} }
\startdata
Resolution &  $45''\times 1.0\mbox{ km s}^{-1}$& $13'' \times 2.03~\mbox{km s}^{-1}$ & $20'' \times 2.6~\mbox{km s}^{-1}$\\

$V_{LSR}$ Coverage (km s$^{-1}$) & $[-340,0]$& $[-340, 0]$ & $[-300, -49]$ \\

Maximum $R_{gal}$ (kpc) & 5.5 & 5.5 & 2.0 \\

Mass Sensitivity\tablenotemark{a} ($M_{\odot}$) & $7.6\times 10^4$ & $5\times 10^4$ & $3.5\times 10^4$ \\

Column Density Sensitivity\tablenotemark{a}(cm$^{-2}$) & $8.2\times 10^{19}$ & $7.5\times10^{20}$ & $2.2\times 10^{20}$ \\

Central Flux\tablenotemark{b} (K km s$^{-1}$) & $1.48\pm 0.02$ & $1.47\pm 0.02$ & $1.50\pm 0.01$ \\

\enddata
\tablenotetext{a}{For a 4$\sigma_{rms}$ detection in two adjacent
channels using a CO-to-H$_2$ conversion factor of $2\times
10^{20}\mbox{ cm}^{-2} (\mbox{K km s}^{-1})^{-1}$.}
\tablenotetext{b}{Flux averaged over the central $7'$ of the galaxy (see
also Figure \ref{datacomp}, {\it right}).}
\end{deluxetable*}

\section{Revisiting the Giant Molecular Clouds of M33}
\label{gmcs-again}


EPRB presented the first complete catalog of GMCs in M33 based on the
BIMA survey data alone.  The $13''$ (50~pc) resolution of the
interferometer data is ideal for identifying individual GMCs though
insufficient for measuring cloud sizes.  The combination of the FCRAO
and BIMA maps is even better for identifying GMCs because the data are
not subject to many of the pathologies associated with
interferometer-only data such as spatial filtering and non-linear flux
recovery.  In addition, the sensitivity of the merged map is
marginally increased over the BIMA data alone because FCRAO and BIMA
sample some of the same spatial frequencies. As shown in
Figure~\ref{comap}, the molecular gas is distributed in many complexes
with low surface brightness filaments connecting the complexes, which
are also connected in velocity space.  The bright CO complex in the
north of the map contains the most massive molecular cloud in the
galaxy (Cloud~1 of EPRB).

\begin{figure}
\plotone{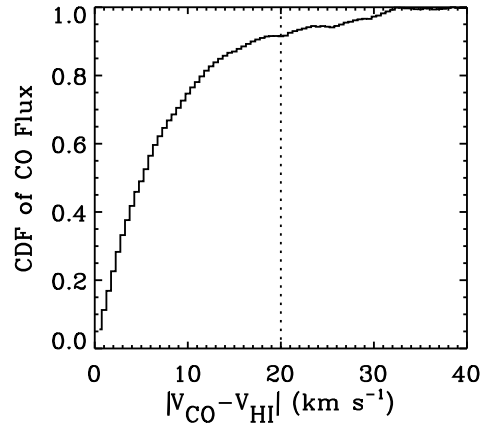}
\caption{\label{cocdf} Cumulative distribution function (CDF) of CO
emission found within a given velocity separation of the \ion{H}{1}
centroid, averaged over the entire CO map. Since 90\% of the CO
emission is found within 20~km~s$^{-1}$ (dotted line), we limit our
search for GMCs over this range. }
\end{figure}

We have generated a new catalog of GMCs in M33 from the merged data
set.  The combined map is less suitable for the identification of GMCs
since the coarser spatial and velocity resolution match GMCs poorly
resulting in a higher completeness limit in the resulting catalog.
The merged catalog is restricted to the region of overlap between the
BIMA and FCRAO maps where all spatial information is recovered.  The
catalog was generated using a contouring method, as used by EPRB. The
first step was to identify all regions in the data set with larger
than a 2.7$\sigma_{rms}$ detection in two adjacent, independent
channels.  For each such detection, all pixels with significance
larger than 2$\sigma_{rms}$ and connected to the 2.7$\sigma_{rms}$
``core'' became part of the cloud candidate.  For each candidate, we
calculated the probability ($P$) of drawing the most significant
spectrum from a random distribution of noise (see Appendix~A of EPRB).
We then multiplied by the number of independent measurements in the
catalog region ($N\approx 2\times 10^{6}$) and considered $-\ln (NP)$
as the statistic of merit.  The final catalog contains all candidates
with $-\ln (NP) > 8.0$ and with CO velocity centroids within
20~km~s$^{-1}$ of the \ion{H}{1} velocity centroid \citep{deul} at
that location.  The selection criteria differ from those of EPRB, who
included high significance detections across the entire bandpass and
found four clouds at large velocity separation from the atomic gas.
Subsequent observations of those apparent clouds showed that they are
not real CO emitters but rather could be attributed to malfunctions in
the BIMA correlator (Rosolowsky, Blitz \& Engargiola, unpublished).
We have selected a 20~km~s$^{-1}$ separation from the \ion{H}{1}
velocity since this velocity range contains 90\% of the CO flux
(Figure \ref{cocdf}) while containing no spurious sources.

Table \ref{catalog} gives the new cloud catalog wherein objects are
named ``M33GMC.''  Clouds are listed in order of increasing
galactocentric radius. The cloud masses are based on a CO-to-H$_2$
conversion factor of $2\times 10^{20}\mbox{ cm}^{-2} (\mbox{K km
s}^{-1})^{-1}$ and the previously-quoted distance of 840~kpc to M33
\citep{Freedman2001}\footnote{For other distances, scale cloud mass in
proportion to $D^2$.}.  The conversion factor is appropriate for the
GMCs in M33 and does not appear to vary significantly over the inner
4~kpc of the galaxy \citep{rpeb03}.  The method for calculating mass
differs from that of EPRB; masses here are based on the method of
\citet{props}.  This method attempts to account for the amount of
emission not included above the 2$\sigma_{rms}$ cutoff in the data
cube by extrapolating the emission profile to the 0~K~km~s$^{-1}$
intensity contour.  Accounting for this emission roughly doubles the
derived cloud masses.  Cloud major and minor axes and orientations are
also given in Table \ref{catalog}; however, the sizes are not
corrected for beam convolution and are thus upper limits.
Deconvolution of the cloud sizes is unstable for large beam sizes (50
pc) and low signal-to-noise values \citep{props}.

\begin{deluxetable*}{cccccccccccc}
\tablecaption{\label{catalog}Catalog of GMC Properties}
\tabletypesize{\footnotesize}
\tablewidth{0pt}
\tablehead{
\colhead{M33GMC} & \colhead{Position} & \colhead{$V_{\mathrm{GMC}}$} &
\colhead{$V_{\mathrm{HI}}$}  & \colhead{$R_{gal}$\tablenotemark{a}} &
\colhead{$V_{\mathrm{GMC}}-V_{\mathrm{HI}}$} & \colhead{$-\ln(NP)$} & 
\colhead{Mass} & \colhead{$\sigma_v$} & \colhead{Size\tablenotemark{b}} & \colhead{Arm} \\
 & \colhead{$(\alpha_{2000},\delta_{2000})$} & 
\colhead{(km s$^{-1}$)} & \colhead{(km s$^{-1}$)} & \colhead{(kpc)} & \colhead{(km s$^{-1}$)} & &
\colhead{($10^5~M_{\odot}$)} & \colhead{(km s$^{-1}$)} &
\colhead{(pc$\times$pc)}& \colhead{Classification}}
\startdata
 1 &  01:33:52.4  +30:39:18 & $-169.0$ & $-169.0$ & 0.18 & $0.0$ & 62.5 & $3.1 \pm 0.9$ & $3.5 \pm 0.9$ & $82 \times 62 ~(122^{\circ})$ & Arm \\
 2 &  01:33:49.5  +30:40:03 & $-191.8$ & $-185.3$ & 0.18 & $-6.6$ & 42.9 & $2.1 \pm 0.5$ & $4.5 \pm 1.9$ & $68 \times 54 ~(12^{\circ})$ & Arm \\
 3 &  01:33:48.1  +30:39:31 & $-168.4$ & $-169.1$ & 0.21 & $0.7$ & 15.1 & $2.5 \pm 2.1$ & $2.6 \pm 2.9$ & $80 \times 62 ~(105^{\circ})$ & Arm \\
 4 &  01:33:48.3  +30:38:53 & $-158.0$ & $-159.7$ & 0.23 & $1.7$ & 32.3 & $1.0 \pm 0.5$ & $2.3 \pm 0.8$ & $62 \times 53 ~(85^{\circ})$ & Arm \\
 5 &  01:33:52.4  +30:40:32 & $-209.9$ & $-201.4$ & 0.24 & $-8.4$ & 34.0 & $1.7 \pm 1.1$ & $2.7 \pm 1.2$ & $67 \times 58 ~(159^{\circ})$ & Arm \\
 6 &  01:33:49.6  +30:41:08 & $-193.6$ & $-199.5$ & 0.46 & $5.9$ & 9.3 & $1.6 \pm 1.5$ & $2.4 \pm 2.0$ & $82 \times 60 ~(36^{\circ})$ & Arm \\
 7 &  01:33:56.8  +30:40:08 & $-192.6$ & $-195.0$ & 0.46 & $2.4$ & 63.9 & $3.5 \pm 1.0$ & $3.0 \pm 0.8$ & $115 \times 63 ~(105^{\circ})$ & Arm \\
 8 &  01:33:44.6  +30:38:54 & $-163.7$ & $-161.0$ & 0.47 & $-2.7$ & 12.3 & $1.4 \pm 1.0$ & $3.1 \pm 3.6$ & $68 \times 58 ~(103^{\circ})$ & Arm \\
 9 &  01:33:51.2  +30:41:29 & $-220.3$ & $-215.4$ & 0.50 & $-4.9$ & 14.2 & $2.5 \pm 1.3$ & $3.6 \pm 1.8$ & $98 \times 50 ~(35^{\circ})$ & Interarm \\
 10 &  01:33:53.5  +30:41:41 & $-224.9$ & $-219.8$ & 0.53 & $-5.1$ & 34.8 & $1.7 \pm 0.7$ & $2.7 \pm 1.1$ & $70 \times 53 ~(34^{\circ})$ & Interarm \\
 11 &  01:33:57.0  +30:41:08 & $-213.4$ & $-211.1$ & 0.54 & $-2.2$ & 20.0 & $1.4 \pm 1.2$ & $2.3 \pm 0.9$ & $68 \times 60 ~(29^{\circ})$ & Interarm \\
 12 &  01:33:55.5  +30:41:37 & $-221.8$ & $-216.4$ & 0.55 & $-5.4$ & 51.0 & $4.8 \pm 1.1$ & $6.5 \pm 1.7$ & $107 \times 82 ~(155^{\circ})$ & Interarm \\
 13 &  01:33:49.9  +30:37:27 & $-149.5$ & $-149.8$ & 0.57 & $0.1$ & 12.4 & $1.1 \pm 1.4$ & $4.9 \pm 3.2$ & $60 \times 48 ~(90^{\circ})$ & Interarm \\
 14 &  01:33:47.9  +30:41:29 & $-203.5$ & $-200.6$ & 0.63 & $-2.9$ & 32.1 & $0.9 \pm 0.5$ & $1.6 \pm 1.4$ & $62 \times 57 ~(70^{\circ})$ & Arm \\
 15 &  01:33:52.5  +30:42:15 & $-214.0$ & $-216.2$ & 0.68 & $2.2$ & 21.7 & $2.6 \pm 1.6$ & $3.7 \pm 2.6$ & $91 \times 52 ~(172^{\circ})$ & Interarm \\
 16 &  01:33:43.1  +30:37:31 & $-139.5$ & $-142.8$ & 0.69 & $3.2$ & 10.1 & $0.4 \pm 0.3$ & $1.8 \pm 2.5$ & $60 \times 49 ~(95^{\circ})$ & Interarm \\
 17 &  01:33:54.8  +30:37:43 & $-167.7$ & $-164.9$ & 0.69 & $-2.8$ & 88.9 & $2.2 \pm 0.4$ & $3.0 \pm 1.0$ & $63 \times 57 ~(89^{\circ})$ & Arm \\
 18 &  01:33:59.9  +30:40:45 & $-209.8$ & $-207.3$ & 0.70 & $-2.4$ & 204.3 & $7.7 \pm 0.9$ & $4.5 \pm 0.6$ & $93 \times 68 ~(107^{\circ})$ & Interarm \\
 19 &  01:33:52.9  +30:37:17 & $-152.7$ & $-154.0$ & 0.71 & $1.3$ & 31.5 & $4.7 \pm 1.4$ & $4.1 \pm 1.3$ & $165 \times 68 ~(1^{\circ})$ & Interarm \\
 20 &  01:33:45.3  +30:36:52 & $-144.2$ & $-139.9$ & 0.73 & $-4.3$ & 30.6 & $1.9 \pm 1.7$ & $2.4 \pm 1.5$ & $66 \times 59 ~(87^{\circ})$ & Interarm \\
 21 &  01:33:59.7  +30:41:32 & $-221.5$ & $-217.3$ & 0.74 & $-4.2$ & 64.2 & $4.5 \pm 1.1$ & $4.9 \pm 1.2$ & $82 \times 64 ~(105^{\circ})$ & Interarm \\
 22 &  01:33:41.0  +30:39:14 & $-166.2$ & $-165.6$ & 0.77 & $-0.6$ & 192.9 & $6.0 \pm 0.7$ & $4.0 \pm 0.9$ & $75 \times 66 ~(128^{\circ})$ & Arm \\
 23 &  01:33:57.8  +30:42:21 & $-233.7$ & $-229.9$ & 0.77 & $-3.8$ & 14.3 & $1.7 \pm 1.0$ & $3.2 \pm 1.6$ & $81 \times 59 ~(165^{\circ})$ & Interarm \\
 24 &  01:33:43.5  +30:41:09 & $-195.0$ & $-195.7$ & 0.83 & $0.7$ & 37.2 & $2.2 \pm 1.0$ & $3.9 \pm 1.3$ & $73 \times 52 ~(135^{\circ})$ & Arm \\
 25 &  01:33:44.3  +30:41:39 & $-198.8$ & $-198.8$ & 0.88 & $-0.1$ & 15.9 & $1.7 \pm 1.4$ & $4.1 \pm 2.2$ & $80 \times 54 ~(76^{\circ})$ & Arm \\
 26 &  01:33:55.8  +30:43:02 & $-227.3$ & $-226.6$ & 0.88 & $-0.7$ & 36.5 & $6.3 \pm 1.6$ & $3.3 \pm 1.2$ & $137 \times 72 ~(7^{\circ})$ & Interarm \\
 27 &  01:33:39.1  +30:38:00 & $-149.4$ & $-148.8$ & 0.91 & $-0.7$ & 17.5 & $1.3 \pm 1.2$ & $3.0 \pm 2.2$ & $73 \times 50 ~(93^{\circ})$ & Arm \\
 28 &  01:33:44.9  +30:36:01 & $-138.7$ & $-138.5$ & 0.93 & $-0.2$ & 61.6 & $3.6 \pm 1.2$ & $4.0 \pm 0.9$ & $70 \times 69 ~(49^{\circ})$ & Interarm \\
 29 &  01:34:02.3  +30:39:12 & $-201.1$ & $-195.4$ & 0.98 & $-5.6$ & 38.3 & $4.2 \pm 1.3$ & $4.1 \pm 2.5$ & $86 \times 67 ~(85^{\circ})$ & Arm \\
 30 &  01:33:57.2  +30:36:41 & $-166.7$ & $-161.9$ & 1.08 & $-4.9$ & 36.3 & $1.1 \pm 0.8$ & $1.7 \pm 0.7$ & $65 \times 61 ~(130^{\circ})$ & Arm \\
 31 &  01:34:02.8  +30:38:38 & $-185.8$ & $-186.4$ & 1.09 & $0.6$ & 33.6 & $5.9 \pm 2.0$ & $6.6 \pm 2.8$ & $173 \times 62 ~(83^{\circ})$ & Arm \\
 32 &  01:33:42.1  +30:35:29 & $-128.0$ & $-129.7$ & 1.11 & $1.5$ & 12.1 & $0.6 \pm 0.5$ & $2.4 \pm 2.3$ & $56 \times 51 ~(179^{\circ})$ & Interarm \\
 33 &  01:33:41.3  +30:41:47 & $-205.6$ & $-196.1$ & 1.11 & $-9.5$ & 16.6 & $2.5 \pm 0.9$ & $3.1 \pm 1.2$ & $79 \times 60 ~(56^{\circ})$ & Arm \\
 34 &  01:33:36.9  +30:39:30 & $-167.7$ & $-163.6$ & 1.13 & $-4.1$ & 43.1 & $6.4 \pm 1.5$ & $3.7 \pm 0.7$ & $138 \times 63 ~(88^{\circ})$ & Arm \\
 35 &  01:33:41.6  +30:35:14 & $-127.2$ & $-130.5$ & 1.17 & $3.2$ & 8.1 & $2.0 \pm 2.5$ & $5.9 \pm 5.0$ & $64 \times 56 ~(136^{\circ})$ & Interarm \\
 36 &  01:34:01.3  +30:43:55 & $-229.2$ & $-228.3$ & 1.20 & $-0.9$ & 44.5 & $3.0 \pm 0.8$ & $2.9 \pm 1.0$ & $69 \times 64 ~(60^{\circ})$ & Interarm \\
 37 &  01:33:35.7  +30:36:25 & $-134.0$ & $-132.8$ & 1.25 & $-1.3$ & 54.4 & $5.0 \pm 1.1$ & $2.7 \pm 0.8$ & $133 \times 72 ~(112^{\circ})$ & Interarm \\
 38 &  01:33:34.8  +30:37:02 & $-136.0$ & $-137.7$ & 1.26 & $1.6$ & 18.9 & $1.5 \pm 0.6$ & $2.3 \pm 0.6$ & $69 \times 59 ~(22^{\circ})$ & Interarm \\
 39 &  01:34:07.2  +30:41:41 & $-217.7$ & $-215.2$ & 1.27 & $-2.5$ & 8.9 & $2.0 \pm 1.2$ & $2.7 \pm 1.4$ & $75 \times 54 ~(73^{\circ})$ & Interarm \\
 40 &  01:34:06.8  +30:39:28 & $-202.5$ & $-200.4$ & 1.32 & $-2.1$ & 38.0 & $1.3 \pm 0.6$ & $4.1 \pm 1.8$ & $62 \times 54 ~(46^{\circ})$ & Interarm \\
 41 &  01:33:59.2  +30:36:08 & $-153.3$ & $-150.7$ & 1.33 & $-2.6$ & 85.3 & $2.8 \pm 0.6$ & $3.8 \pm 1.1$ & $101 \times 57 ~(141^{\circ})$ & Arm \\
 42 &  01:34:01.7  +30:36:43 & $-153.7$ & $-154.7$ & 1.37 & $1.0$ & 40.2 & $3.1 \pm 1.1$ & $2.6 \pm 0.9$ & $79 \times 67 ~(21^{\circ})$ & Arm \\
 43 &  01:33:32.1  +30:36:59 & $-144.7$ & $-141.4$ & 1.45 & $-3.3$ & 14.5 & $0.3 \pm 0.3$ & $1.6 \pm 1.1$ & $49 \times 48 ~(142^{\circ})$ & Interarm \\
 44 &  01:33:31.9  +30:37:58 & $-152.5$ & $-153.0$ & 1.46 & $0.4$ & 9.6 & $1.4 \pm 0.9$ & $3.3 \pm 1.4$ & $75 \times 54 ~(101^{\circ})$ & Interarm \\
 45 &  01:33:48.4  +30:44:44 & $-227.4$ & $-224.7$ & 1.46 & $-2.7$ & 8.2 & $2.1 \pm 1.3$ & $3.3 \pm 1.8$ & $82 \times 64 ~(131^{\circ})$ & Arm \\
 46 &  01:34:08.8  +30:39:11 & $-194.1$ & $-193.2$ & 1.51 & $-0.9$ & 119.5 & $3.4 \pm 0.7$ & $3.0 \pm 0.8$ & $75 \times 61 ~(85^{\circ})$ & Interarm \\
 47 &  01:33:50.4  +30:33:59 & $-134.0$ & $-129.9$ & 1.52 & $-4.2$ & 30.5 & $4.1 \pm 0.9$ & $3.6 \pm 1.1$ & $111 \times 59 ~(54^{\circ})$ & Arm \\
 48 &  01:34:06.3  +30:37:41 & $-163.5$ & $-164.7$ & 1.52 & $1.2$ & 51.8 & $2.9 \pm 0.7$ & $3.5 \pm 1.0$ & $83 \times 62 ~(79^{\circ})$ & Arm \\
 49 &  01:33:33.5  +30:41:22 & $-185.6$ & $-180.3$ & 1.63 & $-5.4$ & 40.9 & $2.5 \pm 0.7$ & $2.6 \pm 1.1$ & $82 \times 57 ~(37^{\circ})$ & Interarm \\
 50 &  01:33:42.8  +30:33:10 & $-119.3$ & $-119.6$ & 1.65 & $0.4$ & 49.9 & $2.7 \pm 1.6$ & $2.8 \pm 1.1$ & $82 \times 68 ~(6^{\circ})$ & Arm \\
 51 &  01:34:11.6  +30:43:15 & $-220.0$ & $-219.7$ & 1.66 & $-0.3$ & 29.9 & $2.3 \pm 0.8$ & $2.8 \pm 1.2$ & $77 \times 62 ~(70^{\circ})$ & Interarm \\
 52 &  01:33:37.4  +30:43:11 & $-201.0$ & $-197.2$ & 1.68 & $-3.8$ & 17.6 & $0.9 \pm 0.6$ & $2.3 \pm 1.1$ & $59 \times 53 ~(164^{\circ})$ & Arm \\
 53 &  01:34:04.0  +30:35:57 & $-146.7$ & $-146.7$ & 1.69 & $-0.1$ & 14.4 & $0.7 \pm 0.5$ & $1.6 \pm 1.3$ & $64 \times 58 ~(87^{\circ})$ & Arm \\
 54 &  01:34:01.1  +30:46:09 & $-238.6$ & $-237.8$ & 1.69 & $-0.7$ & 29.0 & $1.5 \pm 1.2$ & $3.8 \pm 2.3$ & $76 \times 55 ~(30^{\circ})$ & Arm \\
 55 &  01:34:08.6  +30:37:40 & $-156.7$ & $-168.8$ & 1.70 & $12.2$ & 9.7 & $0.7 \pm 1.0$ & $2.3 \pm 1.9$ & $64 \times 53 ~(96^{\circ})$ & Arm \\
 56 &  01:33:59.9  +30:34:49 & $-144.5$ & $-140.5$ & 1.70 & $-3.9$ & 43.3 & $2.4 \pm 1.1$ & $3.5 \pm 1.6$ & $79 \times 61 ~(113^{\circ})$ & Arm \\
 57 &  01:34:12.9  +30:42:00 & $-209.8$ & $-209.5$ & 1.70 & $-0.2$ & 24.0 & $1.8 \pm 1.0$ & $2.7 \pm 1.6$ & $86 \times 52 ~(42^{\circ})$ & Interarm \\
 58 &  01:33:46.7  +30:45:29 & $-232.2$ & $-227.6$ & 1.73 & $-4.5$ & 8.2 & $1.2 \pm 0.9$ & $2.3 \pm 1.8$ & $66 \times 50 ~(46^{\circ})$ & Arm \\
 59 &  01:34:10.8  +30:44:53 & $-234.6$ & $-236.0$ & 1.78 & $1.4$ & 18.9 & $2.7 \pm 1.5$ & $3.2 \pm 2.5$ & $74 \times 66 ~(108^{\circ})$ & Interarm \\
 60 &  01:33:52.4  +30:33:10 & $-127.6$ & $-125.0$ & 1.79 & $-2.5$ & 14.3 & $1.9 \pm 1.4$ & $3.2 \pm 1.4$ & $73 \times 59 ~(79^{\circ})$ & Arm \\
 61 &  01:33:46.9  +30:32:41 & $-117.6$ & $-114.3$ & 1.80 & $-3.3$ & 73.4 & $3.4 \pm 1.2$ & $4.1 \pm 1.4$ & $79 \times 63 ~(29^{\circ})$ & Arm \\
 62 &  01:34:02.6  +30:46:32 & $-245.6$ & $-242.2$ & 1.80 & $-3.4$ & 34.7 & $1.9 \pm 0.6$ & $3.2 \pm 1.2$ & $74 \times 56 ~(51^{\circ})$ & Arm \\
 63 &  01:33:52.5  +30:46:28 & $-240.2$ & $-238.1$ & 1.82 & $-2.1$ & 17.5 & $1.4 \pm 1.3$ & $3.0 \pm 3.2$ & $69 \times 52 ~(173^{\circ})$ & Arm \\
 64 &  01:34:04.3  +30:46:33 & $-248.8$ & $-244.2$ & 1.84 & $-4.6$ & 11.8 & $3.0 \pm 2.0$ & $2.5 \pm 2.8$ & $72 \times 63 ~(121^{\circ})$ & Arm \\
 65 &  01:33:29.0  +30:40:23 & $-179.6$ & $-177.0$ & 1.85 & $-2.6$ & 21.3 & $1.3 \pm 0.9$ & $4.2 \pm 2.6$ & $69 \times 55 ~(95^{\circ})$ & Interarm \\
 66 &  01:34:13.3  +30:39:06 & $-196.2$ & $-196.5$ & 1.89 & $0.3$ & 15.5 & $1.1 \pm 0.9$ & $2.9 \pm 2.4$ & $75 \times 48 ~(141^{\circ})$ & Interarm \\
 67 &  01:33:38.5  +30:32:20 & $-123.6$ & $-119.4$ & 1.89 & $-4.2$ & 13.9 & $0.9 \pm 1.4$ & $2.8 \pm 2.0$ & $66 \times 50 ~(65^{\circ})$ & Arm \\
 68 &  01:33:44.2  +30:32:05 & $-106.3$ & $-111.9$ & 1.93 & $5.6$ & 11.7 & $0.5 \pm 0.7$ & $1.7 \pm 2.4$ & $62 \times 53 ~(179^{\circ})$ & Arm \\
 69 &  01:34:10.8  +30:37:11 & $-168.9$ & $-169.0$ & 1.95 & $0.1$ & 32.8 & $4.1 \pm 1.4$ & $4.4 \pm 1.7$ & $78 \times 61 ~(59^{\circ})$ & Interarm \\
 70 &  01:33:37.9  +30:44:45 & $-211.0$ & $-210.3$ & 2.00 & $-0.7$ & 34.2 & $3.7 \pm 1.8$ & $5.3 \pm 1.6$ & $79 \times 59 ~(128^{\circ})$ & Interarm \\
 71 &  01:33:37.0  +30:31:58 & $-121.5$ & $-116.2$ & 2.01 & $-5.3$ & 54.0 & $0.9 \pm 0.6$ & $1.8 \pm 0.7$ & $65 \times 58 ~(70^{\circ})$ & Arm \\
 72 &  01:33:32.1  +30:32:29 & $-136.7$ & $-126.5$ & 2.03 & $-10.2$ & 16.9 & $0.4 \pm 0.3$ & $2.0 \pm 1.3$ & $56 \times 47 ~(179^{\circ})$ & Arm \\
 73 &  01:33:33.4  +30:32:14 & $-130.4$ & $-122.3$ & 2.03 & $-8.1$ & 25.1 & $1.0 \pm 0.8$ & $2.4 \pm 1.5$ & $71 \times 52 ~(107^{\circ})$ & Arm \\
 74 &  01:34:12.4  +30:37:16 & $-189.0$ & $-179.9$ & 2.06 & $-9.1$ & 44.3 & $1.8 \pm 1.3$ & $4.1 \pm 1.9$ & $66 \times 58 ~(115^{\circ})$ & Interarm \\
 75 &  01:33:33.2  +30:31:56 & $-110.3$ & $-116.9$ & 2.10 & $6.5$ & 30.5 & $0.8 \pm 0.4$ & $1.5 \pm 0.6$ & $62 \times 57 ~(61^{\circ})$ & Arm \\
 76 &  01:34:10.7  +30:36:15 & $-159.2$ & $-164.3$ & 2.11 & $5.0$ & 128.2 & $3.5 \pm 0.5$ & $1.7 \pm 0.7$ & $90 \times 64 ~(168^{\circ})$ & Interarm \\
 77 &  01:34:16.6  +30:39:17 & $-189.5$ & $-191.1$ & 2.13 & $1.6$ & 59.6 & $3.7 \pm 0.9$ & $2.6 \pm 1.3$ & $85 \times 67 ~(47^{\circ})$ & Interarm \\
 78 &  01:33:30.3  +30:32:16 & $-139.9$ & $-125.1$ & 2.14 & $-14.8$ & 17.4 & $1.7 \pm 2.7$ & $2.4 \pm 2.3$ & $126 \times 59 ~(73^{\circ})$ & Arm \\
 79 &  01:33:23.9  +30:39:09 & $-158.9$ & $-160.8$ & 2.15 & $2.0$ & 14.5 & $1.3 \pm 1.0$ & $4.0 \pm 2.3$ & $65 \times 50 ~(133^{\circ})$ & Interarm \\
 80 &  01:33:40.7  +30:46:04 & $-218.6$ & $-217.6$ & 2.17 & $-0.9$ & 88.3 & $8.2 \pm 1.3$ & $5.3 \pm 1.0$ & $145 \times 71 ~(12^{\circ})$ & Interarm \\
 81 &  01:34:06.6  +30:47:52 & $-256.4$ & $-254.2$ & 2.18 & $-2.3$ & 106.6 & $6.2 \pm 0.7$ & $3.5 \pm 0.8$ & $91 \times 68 ~(124^{\circ})$ & Arm \\
 82 &  01:33:29.7  +30:31:56 & $-132.8$ & $-122.9$ & 2.22 & $-10.0$ & 49.4 & $3.8 \pm 1.4$ & $4.7 \pm 1.8$ & $97 \times 71 ~(126^{\circ})$ & Arm \\
 83 &  01:34:03.7  +30:48:16 & $-252.7$ & $-250.0$ & 2.23 & $-2.7$ & 24.8 & $0.6 \pm 0.5$ & $1.7 \pm 0.8$ & $57 \times 53 ~(80^{\circ})$ & Arm \\
 84 &  01:34:15.3  +30:46:27 & $-247.6$ & $-240.1$ & 2.23 & $-7.5$ & 8.4 & $1.0 \pm 1.3$ & $2.5 \pm 1.0$ & $70 \times 54 ~(50^{\circ})$ & Interarm \\
 85 &  01:34:15.8  +30:46:34 & $-236.3$ & $-240.1$ & 2.27 & $3.8$ & 9.7 & $0.9 \pm 0.7$ & $3.1 \pm 2.8$ & $68 \times 50 ~(60^{\circ})$ & Interarm \\
 86 &  01:33:59.1  +30:48:59 & $-248.1$ & $-245.3$ & 2.40 & $-2.8$ & 264.1 & $12.8 \pm 0.8$ & $5.0 \pm 0.7$ & $178 \times 82 ~(21^{\circ})$ & Interarm \\
 87 &  01:34:04.7  +30:49:00 & $-250.2$ & $-249.4$ & 2.42 & $-0.9$ & 17.5 & $2.8 \pm 1.6$ & $3.2 \pm 1.2$ & $91 \times 71 ~(73^{\circ})$ & Interarm \\
 88 &  01:34:08.4  +30:33:51 & $-159.5$ & $-155.9$ & 2.44 & $-3.6$ & 31.6 & $1.6 \pm 0.9$ & $2.6 \pm 1.4$ & $76 \times 53 ~(37^{\circ})$ & Interarm \\
 89 &  01:34:11.7  +30:48:32 & $-259.3$ & $-248.0$ & 2.45 & $-11.3$ & 34.6 & $3.3 \pm 1.2$ & $4.4 \pm 1.4$ & $142 \times 57 ~(137^{\circ})$ & Arm \\
 90 &  01:33:23.9  +30:32:00 & $-122.4$ & $-120.6$ & 2.46 & $-1.8$ & 52.3 & $4.0 \pm 1.3$ & $3.7 \pm 1.0$ & $87 \times 70 ~(101^{\circ})$ & Arm \\
 91 &  01:34:09.2  +30:49:06 & $-249.0$ & $-247.5$ & 2.51 & $-1.5$ & 301.1 & $11.1 \pm 0.8$ & $5.0 \pm 0.8$ & $83 \times 71 ~(61^{\circ})$ & Interarm \\
 92 &  01:33:46.5  +30:48:22 & $-233.2$ & $-234.5$ & 2.51 & $1.3$ & 22.9 & $1.1 \pm 0.8$ & $2.5 \pm 1.2$ & $63 \times 53 ~(157^{\circ})$ & Interarm \\
 93 &  01:34:14.1  +30:48:29 & $-252.8$ & $-249.9$ & 2.52 & $-2.9$ & 14.0 & $0.6 \pm 0.6$ & $1.9 \pm 2.2$ & $57 \times 57 ~(81^{\circ})$ & Arm \\
 94 &  01:34:22.0  +30:39:46 & $-205.2$ & $-204.0$ & 2.53 & $-1.2$ & 8.5 & $0.7 \pm 0.6$ & $4.3 \pm 3.8$ & $82 \times 51 ~(119^{\circ})$ & Interarm \\
 95 &  01:33:42.1  +30:47:46 & $-237.0$ & $-233.6$ & 2.53 & $-3.4$ & 37.5 & $5.0 \pm 1.2$ & $3.4 \pm 1.3$ & $163 \times 66 ~(59^{\circ})$ & Interarm \\
 96 &  01:33:21.3  +30:32:04 & $-120.4$ & $-123.4$ & 2.59 & $2.9$ & 20.5 & $1.9 \pm 1.1$ & $2.5 \pm 1.1$ & $71 \times 61 ~(98^{\circ})$ & Arm \\
 97 &  01:34:14.6  +30:35:11 & $-162.8$ & $-166.8$ & 2.59 & $4.0$ & 9.9 & $0.6 \pm 0.7$ & $2.7 \pm 4.0$ & $63 \times 47 ~(99^{\circ})$ & Interarm \\
 98 &  01:33:38.9  +30:47:23 & $-225.1$ & $-225.3$ & 2.59 & $0.2$ & 10.2 & $0.8 \pm 1.0$ & $3.0 \pm 2.3$ & $60 \times 57 ~(98^{\circ})$ & Interarm \\
 99 &  01:34:13.8  +30:34:38 & $-174.1$ & $-171.3$ & 2.65 & $-2.8$ & 43.3 & $4.4 \pm 1.4$ & $4.5 \pm 1.2$ & $89 \times 72 ~(162^{\circ})$ & Interarm \\
 100 &  01:34:08.1  +30:32:50 & $-145.3$ & $-149.5$ & 2.66 & $4.2$ & 16.0 & $2.3 \pm 1.5$ & $2.6 \pm 1.6$ & $110 \times 59 ~(29^{\circ})$ & Interarm \\
 101 &  01:34:17.7  +30:48:36 & $-248.5$ & $-245.5$ & 2.68 & $-3.0$ & 24.8 & $1.2 \pm 1.0$ & $2.2 \pm 1.7$ & $67 \times 51 ~(66^{\circ})$ & Arm \\
 102 &  01:34:18.5  +30:48:23 & $-245.1$ & $-244.2$ & 2.68 & $-0.9$ & 14.0 & $1.2 \pm 0.8$ & $3.5 \pm 1.8$ & $64 \times 49 ~(29^{\circ})$ & Arm \\
 103 &  01:34:10.0  +30:49:53 & $-261.1$ & $-258.6$ & 2.70 & $-2.4$ & 10.3 & $1.8 \pm 0.6$ & $3.5 \pm 1.9$ & $97 \times 54 ~(92^{\circ})$ & Interarm \\
 104 &  01:33:18.5  +30:40:56 & $-178.8$ & $-172.7$ & 2.77 & $-6.1$ & 15.2 & $1.1 \pm 0.9$ & $2.3 \pm 1.6$ & $67 \times 63 ~(164^{\circ})$ & Interarm \\
 105 &  01:34:13.6  +30:33:44 & $-156.6$ & $-157.9$ & 2.82 & $1.3$ & 158.2 & $5.0 \pm 0.7$ & $4.0 \pm 0.9$ & $70 \times 64 ~(30^{\circ})$ & Interarm \\
 106 &  01:34:19.2  +30:49:05 & $-227.5$ & $-246.1$ & 2.83 & $18.6$ & 10.0 & $0.7 \pm 0.7$ & $3.1 \pm 2.2$ & $58 \times 50 ~(177^{\circ})$ & Arm \\
 107 &  01:33:18.5  +30:41:34 & $-178.6$ & $-175.5$ & 2.85 & $-3.1$ & 31.7 & $1.5 \pm 0.8$ & $3.0 \pm 1.5$ & $69 \times 57 ~(171^{\circ})$ & Interarm \\
 108 &  01:33:39.3  +30:28:20 & $-104.3$ & $-104.9$ & 2.88 & $0.6$ & 8.9 & $0.8 \pm 1.0$ & $2.0 \pm 1.4$ & $79 \times 49 ~(59^{\circ})$ & Interarm \\
 109 &  01:34:22.6  +30:48:45 & $-249.5$ & $-244.2$ & 2.93 & $-5.3$ & 23.6 & $5.5 \pm 1.4$ & $3.1 \pm 1.2$ & $138 \times 68 ~(88^{\circ})$ & Arm \\
 110 &  01:34:10.0  +30:31:52 & $-148.7$ & $-148.5$ & 3.01 & $-0.1$ & 10.9 & $0.6 \pm 0.8$ & $2.3 \pm 1.5$ & $72 \times 47 ~(9^{\circ})$ & Interarm \\
 111 &  01:33:13.6  +30:39:28 & $-160.6$ & $-160.0$ & 3.02 & $-0.7$ & 38.0 & $1.3 \pm 0.7$ & $2.8 \pm 2.2$ & $59 \times 59 ~(171^{\circ})$ & Interarm \\
 112 &  01:33:54.0  +30:51:04 & $-247.5$ & $-255.8$ & 3.03 & $8.4$ & 10.6 & $1.4 \pm 1.8$ & $2.9 \pm 3.8$ & $61 \times 53 ~(49^{\circ})$ & Interarm \\
 113 &  01:33:32.0  +30:47:37 & $-214.4$ & $-214.0$ & 3.04 & $-0.4$ & 25.5 & $0.9 \pm 0.6$ & $3.5 \pm 2.2$ & $52 \times 51 ~(144^{\circ})$ & Interarm \\
 114 &  01:33:15.3  +30:41:14 & $-165.0$ & $-166.7$ & 3.07 & $1.7$ & 24.2 & $1.3 \pm 0.5$ & $3.1 \pm 1.7$ & $74 \times 54 ~(62^{\circ})$ & Interarm \\
 115 &  01:33:18.4  +30:29:38 & $-106.5$ & $-102.4$ & 3.10 & $-4.0$ & 11.9 & $0.5 \pm 0.7$ & $1.9 \pm 1.6$ & $68 \times 56 ~(92^{\circ})$ & Interarm \\
 116 &  01:34:17.7  +30:33:41 & $-159.3$ & $-157.3$ & 3.11 & $-2.0$ & 21.2 & $2.2 \pm 2.4$ & $3.6 \pm 2.0$ & $106 \times 67 ~(132^{\circ})$ & Interarm \\
 117 &  01:33:12.0  +30:39:32 & $-162.7$ & $-159.8$ & 3.15 & $-2.9$ & 21.6 & $1.3 \pm 0.9$ & $3.8 \pm 1.5$ & $84 \times 48 ~(27^{\circ})$ & Interarm \\
 118 &  01:34:18.7  +30:33:48 & $-166.7$ & $-158.8$ & 3.17 & $-7.9$ & 10.0 & $0.6 \pm 1.1$ & $2.3 \pm 2.1$ & $61 \times 54 ~(54^{\circ})$ & Interarm \\
 119 &  01:34:21.7  +30:34:55 & $-187.6$ & $-179.0$ & 3.18 & $-8.7$ & 12.7 & $0.7 \pm 0.7$ & $3.6 \pm 1.8$ & $88 \times 47 ~(90^{\circ})$ & Interarm \\
 120 &  01:34:27.4  +30:49:12 & $-248.4$ & $-245.0$ & 3.24 & $-3.5$ & 21.6 & $4.0 \pm 1.1$ & $3.7 \pm 1.2$ & $111 \times 61 ~(51^{\circ})$ & Interarm \\
 121 &  01:34:16.7  +30:51:39 & $-266.9$ & $-264.3$ & 3.25 & $-2.6$ & 10.3 & $1.9 \pm 1.1$ & $3.1 \pm 2.0$ & $80 \times 60 ~(25^{\circ})$ & Interarm \\
 122 &  01:34:31.7  +30:47:37 & $-252.5$ & $-246.2$ & 3.32 & $-6.3$ & 16.1 & $1.7 \pm 1.2$ & $2.5 \pm 1.6$ & $96 \times 64 ~(61^{\circ})$ & Arm \\
 123 &  01:34:31.9  +30:47:42 & $-243.2$ & $-246.1$ & 3.34 & $2.8$ & 11.3 & $2.3 \pm 1.2$ & $3.1 \pm 1.3$ & $66 \times 66 ~(12^{\circ})$ & Arm \\
 124 &  01:34:33.0  +30:46:47 & $-243.0$ & $-242.0$ & 3.35 & $-1.1$ & 125.2 & $10.0 \pm 1.1$ & $4.1 \pm 0.8$ & $117 \times 87 ~(149^{\circ})$ & Arm \\
 125 &  01:34:15.9  +30:52:19 & $-266.1$ & $-262.0$ & 3.37 & $-4.1$ & 11.8 & $2.3 \pm 1.4$ & $3.3 \pm 1.6$ & $87 \times 55 ~(79^{\circ})$ & Interarm \\
 126 &  01:34:34.5  +30:46:19 & $-221.8$ & $-235.1$ & 3.42 & $13.2$ & 96.0 & $3.5 \pm 0.7$ & $3.9 \pm 1.0$ & $78 \times 60 ~(27^{\circ})$ & Arm \\
 127 &  01:33:05.7  +30:35:08 & $-144.4$ & $-144.0$ & 3.47 & $-0.4$ & 31.4 & $0.9 \pm 0.6$ & $2.0 \pm 1.1$ & $62 \times 61 ~(86^{\circ})$ & Interarm \\
 128 &  01:33:44.0  +30:51:47 & $-241.5$ & $-236.9$ & 3.52 & $-4.6$ & 16.6 & $1.9 \pm 1.8$ & $4.6 \pm 5.2$ & $61 \times 54 ~(56^{\circ})$ & Interarm \\
 129 &  01:33:39.9  +30:51:22 & $-251.0$ & $-233.2$ & 3.58 & $-17.8$ & 8.4 & $0.6 \pm 0.5$ & $2.7 \pm 1.8$ & $57 \times 48 ~(172^{\circ})$ & Interarm \\
 130 &  01:34:14.4  +30:30:26 & $-151.8$ & $-150.7$ & 3.61 & $-1.2$ & 16.3 & $0.6 \pm 0.5$ & $2.3 \pm 1.3$ & $82 \times 57 ~(9^{\circ})$ & Interarm \\
 131 &  01:34:22.8  +30:32:41 & $-166.0$ & $-158.1$ & 3.68 & $-7.9$ & 33.0 & $2.4 \pm 1.0$ & $2.7 \pm 1.1$ & $91 \times 63 ~(92^{\circ})$ & Interarm \\
 132 &  01:33:23.6  +30:25:40 & $-112.2$ & $-109.8$ & 3.70 & $-2.5$ & 78.9 & $4.4 \pm 0.7$ & $3.1 \pm 0.8$ & $82 \times 69 ~(121^{\circ})$ & Interarm \\
 133 &  01:34:39.0  +30:43:57 & $-220.1$ & $-217.6$ & 3.71 & $-2.5$ & 11.8 & $0.5 \pm 0.4$ & $1.3 \pm 1.4$ & $61 \times 54 ~(63^{\circ})$ & Arm \\
 134 &  01:33:13.0  +30:27:22 & $-115.4$ & $-119.7$ & 3.71 & $4.2$ & 19.7 & $2.7 \pm 1.7$ & $3.0 \pm 1.1$ & $116 \times 60 ~(2^{\circ})$ & Interarm \\
 135 &  01:33:10.6  +30:27:43 & $-122.5$ & $-117.0$ & 3.76 & $-5.6$ & 28.5 & $2.2 \pm 0.8$ & $2.8 \pm 1.1$ & $82 \times 54 ~(39^{\circ})$ & Interarm \\
 136 &  01:34:38.3  +30:40:30 & $-199.9$ & $-202.6$ & 3.80 & $2.7$ & 108.1 & $10.7 \pm 1.1$ & $8.4 \pm 1.5$ & $239 \times 106 ~(52^{\circ})$ & Arm \\
 137 &  01:33:03.7  +30:39:56 & $-164.4$ & $-165.2$ & 3.87 & $0.7$ & 16.9 & $0.8 \pm 0.8$ & $3.2 \pm 1.4$ & $55 \times 51 ~(144^{\circ})$ & Interarm \\
 138 &  01:34:41.4  +30:47:10 & $-236.3$ & $-238.6$ & 3.95 & $2.4$ & 12.4 & $1.4 \pm 0.9$ & $2.8 \pm 2.6$ & $74 \times 55 ~(112^{\circ})$ & Interarm \\
 139 &  01:34:04.1  +30:55:09 & $-256.5$ & $-256.6$ & 3.98 & $0.0$ & 11.8 & $1.9 \pm 1.0$ & $4.2 \pm 2.6$ & $75 \times 53 ~(37^{\circ})$ & Interarm \\
 140 &  01:34:33.4  +30:35:17 & $-176.5$ & $-172.7$ & 4.02 & $-3.7$ & 24.0 & $1.6 \pm 0.9$ & $2.1 \pm 1.1$ & $115 \times 57 ~(132^{\circ})$ & Interarm \\
 141 &  01:34:36.0  +30:36:22 & $-182.7$ & $-179.8$ & 4.06 & $-2.9$ & 8.1 & $0.9 \pm 1.1$ & $5.3 \pm 4.3$ & $74 \times 53 ~(146^{\circ})$ & Interarm \\
 142 &  01:34:29.7  +30:55:30 & $-253.3$ & $-262.9$ & 4.42 & $9.6$ & 10.3 & $1.6 \pm 0.4$ & $3.8 \pm 1.3$ & $62 \times 60 ~(36^{\circ})$ & Interarm \\
 143 &  01:34:07.2  +30:57:52 & $-266.8$ & $-258.6$ & 4.68 & $-8.1$ & 8.4 & $1.3 \pm 1.1$ & $2.3 \pm 2.9$ & $96 \times 56 ~(45^{\circ})$ & Interarm \\
 144 &  01:32:44.2  +30:35:12 & $-144.0$ & $-148.0$ & 5.17 & $4.0$ & 10.3 & $1.2 \pm 0.6$ & $4.0 \pm 2.3$ & $69 \times 51 ~(55^{\circ})$ & Interarm \\
 145 &  01:33:16.9  +30:52:53 & $-219.7$ & $-216.6$ & 5.21 & $-3.1$ & 9.2 & $0.9 \pm 0.4$ & $3.6 \pm 2.1$ & $61 \times 48 ~(50^{\circ})$ & Interarm \\
 146 &  01:32:42.1  +30:36:35 & $-155.4$ & $-153.8$ & 5.39 & $-1.6$ & 8.5 & $0.9 \pm 0.5$ & $2.1 \pm 1.1$ & $58 \times 52 ~(166^{\circ})$ & Interarm \\
 147 &  01:32:42.2  +30:22:27 & $-101.6$ & $-101.4$ & 5.98 & $-0.1$ & 12.4 & $1.1 \pm 0.6$ & $2.5 \pm 1.2$ & $62 \times 55 ~(92^{\circ})$ & Interarm \\
 148 &  01:33:10.5  +30:56:42 & $-202.4$ & $-214.4$ & 6.50 & $12.0$ & 8.6 & $1.0 \pm 0.9$ & $6.9 \pm 3.5$ & $59 \times 49 ~(175^{\circ})$ & Interarm \\
 149 &  01:32:11.5  +30:30:18 & $-134.2$ & $-128.5$ & 7.65 & $-5.7$ & 8.4 & $0.9 \pm 0.9$ & $3.3 \pm 2.5$ & $58 \times 51 ~(19^{\circ})$ & Interarm \\

\enddata
\tablenotetext{a}{Derived assuming $D=840\mbox{kpc}$ 
\citep{Freedman2001}, $i=52^{\circ}$ and PA$=22^{\circ}$ \citep{cs00}.}
\tablenotetext{b}{Non-deconvolved major and minor axes of clouds,
extrapolated to the zero intensity contour (with position angle).}
\end{deluxetable*}

It is possible that the catalog method adopted in this study will
blend together individual GMCs into a larger complex that, at higher
resolution, would be identified as separate clouds.  In their study of
this problem, EPRB found many instances of cloud groups identified as
distinct in the high-resolution work of \citet{ws90} that were merged
in the EPRB catalog. The same effects must be present in the current
catalog and will affect studies of the mass distributions below
\S\ref{massspec}.  In general, appropriately decomposing a blend moves
an individual cloud to multiple clouds of lower masses, steepening the
mass distribution and reinforcing any truncations in the distribution
(provided the mass of the original cloud was near the truncation
mass).  We proceed with this caveat in mind but make no effort to
separate the catalog objects into smaller units.  Decomposing blends
is inaccurate at this physical resolution \citep[50 pc;][]{props}.

The new catalog contains 149 clouds to an estimated completeness limit
of $1.3\times 10^{5}$~\Msun.  The limit is based on linearly scaling
the completeness limit derived from EPRB's detailed examination of the
BIMA-only survey to the sensitivity limit in the BIMA+FCRAO data. The
noise level of the merged map does not vary significantly in the
region where molecular clouds are found, although it does increase in
the outer regions of the galaxy.  The original catalog of EPRB
contained 144 clouds (omitting the spurious high-velocity outliers)
with total mass $1.7\times 10^7$~\Msun.  Total mass of clouds in the
new catalog would be $2.0\times 10^7$~\Msun\ using the same mass
formula as EPRB.  This increase in the mass estimate stems from
inclusion of the low surface brightness emission that is not detected
in the interferometer-only map.  The larger increase in the catalog
GMC mass, to $3.9\times 10^7$~\Msun, comes from accounting for the
flux below the 2$\sigma_{rms}$ limit.  This total mass accounts for
about 1/3 of all the mass implied by emission in the merged map,
$1.2\times 10^8$~\Msun.  (This differs from the value reported by HCSY
in part because we include helium and adopt a different distance and
CO-to-H$_2$ conversion factor.)  The remaining 2/3 of the emission is
from sources that are of smaller mass than GMCs.  This represents
quite a contrast to the inner Milky Way, where the GMCs comprise 80\%
of the molecular mass \citep{wm97}. 

\begin{figure}
\plotone{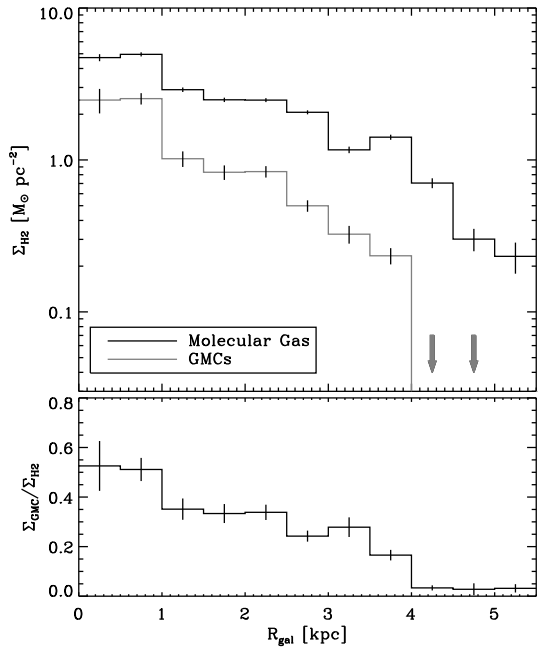}
\caption{\label{sdcompare} Radial surface density of molecular gas
mass.  The top panel shows the total mass surface density (upper
curve) and the surface density of mass in GMCs ($M>10^5~M_{\odot}$ ---
lower curve).  The bottom panel shows the ratio between these
quantities.  Error bars show the uncertainties in each 500~pc bin. At
nearly all radii, GMCs comprise a minority of the total molecular gas
mass, and the fraction declines from the galactic center. There
appears to be a sharp cutoff in the distribution of GMCs at
$R_{gal}\sim 4.0$~kpc, but the total molecular gas surface density
profile continues beyond that radius.}
\end{figure}

The amount of CO emission from low mass clouds can be examined by
averaging the data in annuli of galactocentric
radius. Figure~\ref{sdcompare} compares the radial surface density
profile for GMCs ($M>10^5~M_{\odot}$) and for all CO emission.  Within
1~kpc of M33's center, GMCs comprise $\lesssim$60\% of the total
molecular mass, and the fraction decreases rapidly with radius to
$<$10\% near $R_{gal}\approx 4.0$~kpc.  Beyond that radius, GMCs
vanish almost entirely with only a few, low-mass clouds beyond this
radius although there is plenty of molecular gas.  The dearth of GMCs
beyond 4~kpc is real: the catalog completeness limit is only 10\%
higher in mass at $R_{gal}=5$~kpc than at $R_{gal}=0$. Thus there are
significantly fewer GMCs per unit total molecular mass beyond
$R_{gal}=4.1$~kpc than within that radius.  The radius 4~kpc
corresponds to the outer ends of the major spiral arms in M33
identified in \ion{N}{1} and \ion{S}{1} \citep{hs80}.  In addition,
the neon abundance shows a sharp drop at this radius \citep{m33-neon}.
However, star formation as traced by H$\alpha$ continues out to
$R_{gal}=6.7$~kpc \citep{k89}, and the disk is gravitationally
unstable to that radius \citep{c03}.  Whatever the explanation for the
changing conditions at $R_{gal}>4$~kpc, star formation in the outer
part of M33 must occur in molecular clouds with
$M\la1.3\times10^5$~\Msun.  It is worth noting that this effect
confounded the total mass estimate of EPRB who extrapolated the GMC
mass to the total mass of the galaxy based on the fraction of emission
in GMCs for the center of the galaxy.

\begin{figure}
\plotone{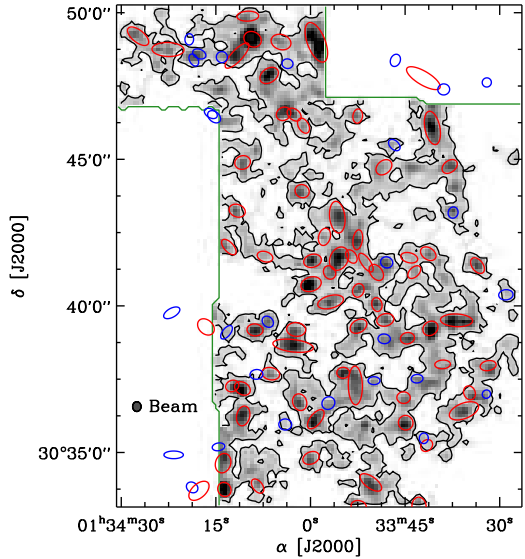}
\caption{\label{deepmap} Map of the integrated intensity of CO
emission in the central region of M33 in gray scale from the combined
NRO+BIMA+FCRAO map.  The gray scale runs linearly from 0 to 8
K~km~s$^{-1}$ and a contour is drawn at 1 K~km~s$^{-1}$.  Positions of
GMCs are indicated with red (blue) ellipses for clouds above (below)
the $1.3\times 10^5~M_{\odot}$ completeness limit of the revised
catalog from the BIMA+FCRAO map.  The extent of the ellipses is
determined by the size of the object in the merged map, extrapolated
to the zero intensity contour and convolved to the resolution of the
combined map.  The border of the NRO observations is shown as a green
line.  Faint CO emission associated with low mass clouds is found
around the high mass molecular clouds.}
\end{figure}

\begin{figure}
\plotone{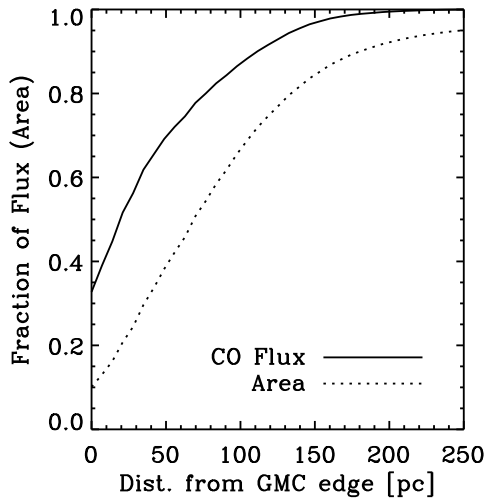}
\caption{\label{localized} Cumulative distribution of CO flux as a
function of separation from the edges of the GMCs shown in Figure
\ref{deepmap} using the combined BIMA+FCRAO+NRO map. The solid line
represents the flux distribution and the dotted line shows the
cumulative distribution of the survey area.  The excess of the flux
curve over the area curve shows the spatial correlation of the CO flux
from low-mass clouds with the cataloged GMCs.  Since the separation is
measured from the {\it edge} of GMCs rather than the center, the
curves begin with the 33\% of the emission found in GMCs which are found
in the 10\% of the map area.  }
\end{figure}

The location of the low mass clouds in the inner region of M33 is
shown directly in the combined map (Figure~\ref{deepmap}) with its
50\% lower rms noise.  The size of objects identified in the merged
map is indicated with ellipses, representing the size and orientation
of the objects as they would appear in the combined map after
accounting for the differing resolutions and sensitivities.  Emission
beyond the boundaries of the identified GMCs arises from low mass
clouds.  This emission forms a filamentary structure, generally
surrounding and linking the high mass GMCs. The GMCs themselves appear
to define the compact peaks of the lower surface brightness
structures.  The low mass clouds are clearly correlated with the high
mass GMCs, and we quantify this relationship by measuring the amount
of flux found within a given separation from {\it the edge} of a
catalog GMC.  The results are shown in Figure \ref{localized} which
plots the cumulative distribution of CO flux as a function of
projected separation from the edge of the catalog GMCs shown in
Figure~\ref{deepmap}.  The curve rises from $\sim 30\%$ at zero
separation (the amount of flux in GMCs) to over 90\% at 100 pc
separation.  The dotted curve indicates the fraction of the area found
within the same separation, which would be the curve that the flux
distribution would follow if there were no spatial correlation between
low mass and high mass clouds.  The excess above the dotted curve
shows the spatial correlation of the emission. There is no evidence
for a galaxy-spanning, diffuse molecular gas component, confirming the
conclusions of \citet{rpeb03}.

\section{Variations in the GMC Mass Distribution}
\label{massspec}

\subsection{Radial Variation}
\label{radvar}
The cloud size distribution varies with galactocentric radius in a
surprising way.  Fig.~\ref{sdcompare} shows the absence of GMCs 
beyond 4.1~kpc, but the most massive clouds ($M>8\times10^5$~\Msun) are
also absent inside a radius of 2.1~kpc.  Indeed all six clouds with
$M>8\times10^5$~\Msun\ are at $2.1\la R_{gal}\la2.5$~kpc.  Evidently
something in the inner galaxy eliminates very large clouds, and
something in the outer galaxy eliminates  GMCs nearly altogether.

\begin{figure}
\plotone{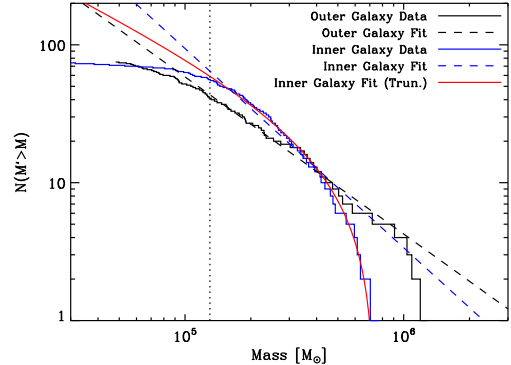}
\caption{\label{inout} Cumulative mass distribution of GMCs in the inner
(blue, $R_{gal}<2.1$ kpc) and outer (black, $R_{gal}>2.1$ kpc) regions
of M33.  A power law cumulative mass distribution is shown for the
outer galaxy (black dashed line), and truncated and non-truncated
power law distributions are shown for the inner galaxy (blue dashed
and red solid lines respectively).  The vertical dotted line indicates
the completeness limit for the survey.}
\end{figure}

To examine the cloud distribution more quantitatively, we consider
two regions divided at $R_{gal}= 2.1$~kpc.  This division yields
approximately equal GMC mass in each region.  Figure~\ref{inout}
shows the mass functions for the GMCs in these two regions.  The mass
distributions are significantly different.  In the outer GMC region
(the annulus $2.1 < R_{gal} < 4.1$~kpc), the GMC mass distribution
follows a power law from high mass down to the completeness limit of
the catalog.  In the inner region, however, despite the existence of
many more GMCs (about three times more per unit area), the mass
distribution is truncated at the high end and may also fall slightly
below a power law distribution at the low end.  A two-sided KS test
\citep{numrec} shows there is only a 0.8\% probability that the two
samples could be drawn from a single distribution while exhibiting
such a large difference.

The cumulative mass distribution functions can be characterized
quantitatively by truncated power-law functions with the form:
\begin{equation}
N(M'>M)= N_0 \left[\left(\frac{M}{M_0}\right)^{\gamma+1}-1\right],
\label{cumdist}
\end{equation}
where $N_0$, $M_0$, and $\gamma$ are parameters, and the fit is
restricted to clouds above the completeness limit of $1.3\times
10^5~M_{\odot}$.  This functional form is adopted for
comparison with the GMC population in the Milky Way.  The parameter
$\gamma$ is the index of the mass 
distribution.  If $\gamma<-2$,  most of the mass is found in low
mass clouds, and if $\gamma>-2$,  the high mass clouds dominate as
is the case for the inner Milky Way.  As long as $N_0$ is significantly larger
than unity, the mass distribution has a truncation near 
$M=2^{1/(\gamma+1)} M_0$.  In the inner
Milky Way, the mass distribution is commonly expressed as a
differential mass distribution.  The differential mass distribution is
well-characterized by a power-law with a truncation at large cloud
masses \citep[$M_0\sim 6\times 10^6~M_{\odot}$][and references
therein]{wm97}:
\begin{equation}
\label{diff}
\frac{dN}{dM}\propto \left(\frac{M}{M_0}\right)^{\gamma},~
M<M_0.
\end{equation}
A similar form of the distribution, albeit with different slopes, has
been reported in the outer Milky Way \citep[e.g.][]{hc01}, the LMC
\citep{nanten-mspec}, M33 (EPRB), and other systems farther afield.
Fitting the cumulative distribution instead of the differential one is
a better way to describe the data because the cumulative function is
not affected by biases introduced by binning, and it can account for
uncertainties in the cloud masses
\citep{mspec}. Table~\ref{mspectable} gives the best-fit coefficients
for the M33 data, and the resulting distributions are plotted
Figure~\ref{inout}.  As noted above, there is no significant
truncation of the mass distribution for the outer galaxy clouds, and
Table~\ref{mspectable} reports coefficients for a simple power law
fit.  For comparison, Table~\ref{mspectable} also includes the fit to
the cumulative mass distribution of all the clouds in the catalog.
Both the fits and Figure~\ref{inout} indicate that the primary
difference between the two distributions is the truncation in the
upper end of the mass distribution of the inner galaxy clouds.  No
such truncation is apparent in the outer portion of the galaxy, and
including one in the fit does not appreciably change the derived
values or goodness of fit.  In the mass range where the inner galaxy
distribution fits a power law, $1.3 < M/(10^5~M_{\odot})< 5.0$, the
two distributions are indistinguishable.  While the inner galaxy has a
larger fraction of its molecular gas in GMCs (Figure~\ref{sdcompare}),
the formation or survival of the highest-mass GMCs ($\gtrsim 7\times
10^{5}~M_{\odot}$) is suppressed.  One possibility is that high mass
clouds are sheared apart by galactic tides in this region; but this
possibility is unlikely because clouds with masses of $10^7~M_{\odot}$
would be stable\footnote{We have assumed the surface densities and
virial parameters would comparable to other GMCs in M33 and used
average values of these quantities from \citet{psp5}.}  according to
the criteria of \citet{tides}.

\begin{deluxetable}{ccccc}
\tablecaption{\label{mspectable}Parameters of Mass Distributions in M33}
\tablewidth{0pt}
\tablehead{
\colhead{Objects} & \colhead{$\gamma$} & \colhead{$N_0$} & 
\colhead{$M_0/(10^5~M_\odot)$}}
\startdata
All Clouds & $-2.0 \pm 0.2$ & $11 \pm 2$ & $ 14 \pm 7$ \\
\cutinhead{Radial Variation}
Outer galaxy & $-2.1\pm 0.1$ &  \nodata  & $34\pm 14$  \\
Inner galaxy & $-1.8 \pm 0.2$ & $21 \pm 11$ & $7.4 \pm 0.5$ \\
Inner galaxy\tablenotemark{a} & $-2.4\pm 0.2$ & \nodata & $23 \pm 5$  \\
\cutinhead{North-South Variation}
North & $-2.0\pm 0.2$ & $5.4\pm 4.3$ & $34\pm 12$  \\
North & $-2.25\pm 0.15$ & \nodata & $34\pm 11$ \\
South & $-1.4 \pm 0.2$ & $41 \pm 11$ & $6.4\pm 0.5$ \\
South\tablenotemark{a} & $-2.4 \pm 0.3$ & \nodata & $20\pm 16$ \\
\cutinhead{Arm-Interarm Variation}
Arm & $-1.9 \pm 0.2$ & $5 \pm 4$ & $16 \pm 5$ \\
Interarm & $-2.2 \pm 0.2$ & $4 \pm 4$ & $13 \pm 4$ \\	
\enddata

\tablenotetext{a}{A mass spectrum without a truncation is also fit to
the data as shown in Figure \ref{inout} to illustrate the
significance of the cutoff found in the southern/inner galaxy clouds.}
\end{deluxetable}

The mass truncation and differences in the slope of the distribution
may be effected by several aspects of the observations and the
cataloging procedure, but none of the likely effects would produce a
truncation where none is present.  \citet{ws90} have already reported
a truncation in the mass distribution of GMCs, but it is difficult to
assess their results as their survey is spatially incomplete and
selected targets based on dust and CO priors.  Since the merged data
span the entire galaxy, spatial biases will not affect the mass
distribution.  The \citet{ws90} data had significantly higher
resolution and several of the clouds considered as separate in their
study are not resolved into individual clouds here (or in EPRB, see
their \S 3.5 and Figures 13 and 14).  Blending will act to minimize
the effects of truncation since separate small clouds will appear as a
single cloud of higher mass, so the truncations are certainly present
in the GMC mass distributions.  The correction for emission found
below the $2\sigma$ boundary of the clouds will also change the mass
distribution slightly.  Since flux loss primarily affects clouds near
the completeness limit (since most of the emission in these clouds is
below the identification contour), correcting for the effects of the
flux loss will move these clouds to higher mass, thereby steepening
the spectrum.  Since this is a correction for an instrumental effect,
making this correction likely improves estimates of the mass
distribution.  Finally, regardless of the precise nature of the
systematic effects on the measured mass distributions, this reported
measurement is {\it differential} between two regions cataloged with a
uniform method.  As such, the reported differences must reflect some
underlying difference in the character of molecular gas in the two
regions.

Similar variations in the GMC mass distribution as a function of
position in a galaxy have not been directly quantified in previous
work, but there is some evidence for such variations in the Milky Way.
\citet{mspec} argued for different mass distribution functions between the
inner and outer Galaxy.  This variation within the Milky Way is
uncertain, however, because differences in cloud cataloging methods
can produce apparent differences in the mass distribution while the
underlying cloud populations are the same.  In addition, large
arm-interarm contrasts observed in many spiral galaxies suggest that
the mass distribution of GMCs changes azimuthally (see also
\S\ref{sparm}).


\subsection{North-South Variation}
\label{northsouth}
The right-hand panel of Figure \ref{datacomp} implies that the
approaching and receding sides of the galaxy have a significant flux
asymmetry.  The northern (approaching) portion of the galaxy has a total
flux of $8200\mbox{ Jy km s}^{-1}$ whereas the southern (receding)
half of the galaxy has a flux of $7800\mbox{ Jy km s}^{-1}$, a
difference of $5\%$.  The asymmetry is also seen in the \ion{H}{1}
(where the approaching half of the galaxy has 13\% more flux than the
receding) and the H$\alpha$ (5\%) distributions \citep[based on the
images of][]{lgs-m31m33}, making M33 slightly lopsided.  Motivated by
this flux asymmetry, we also consider the variation in the GMC mass
distributions between the northern and southern halves of the galaxy.
We define the southern (receding) portion of the galaxy as the
portion of the disk with galactocentric azimuth within $90^{\circ}$ of
the kinematic major axis and the northern region as the complement of
the southern region.  We fit truncated power-laws to the two
distributions and report the derived parameters in Table
\ref{mspectable}.  Like the variation between the inner and the outer
galaxies, there is substantial difference between the northern and
southern mass distribution, particularly at high masses.  The southern
portion of the galaxy lacks high mass clouds that are present in the
northern portion of the galaxy.  Both mass distributions have some
evidence for truncation at the high masses, though the evidence is
relatively weak for clouds in the northern portion of the galaxy.
While it would be interesting to separate the variation in the mass
distributions into 4 separate regions (inner vs. outer and northern
vs. southern), we lack sufficient numbers of clouds for reliable fits.
Since differences between the inner and outer regions of the galaxy
are more pronounced, we focus our analysis on these difference, though
with better data, the north-south asymmetry could be explored.

\subsection{Arm-Interarm Variation}
\label{sparm}
For galaxies with well defined spiral structure, the spiral arms
contain nearly all of the molecular material in the galaxy
\citep{m51-co,song,iram-m31-aa}.  The Milky Way appears to be such a
galaxy \citep[e.g.][]{dht01,mw-spiral}.  Molecular gas in the interarm
regions of spiral galaxies appears to have different properties than
that found in the arms \citep{n6946-walsh,m51-1213,mw-spiral} though
\citet{gmas-5055} suggest that the properties of individual GMCs may
not vary significantly.  While our observations lack the resolution to
assess whether the GMC properties are substantially different in
interarm regions of M33, we can investigate whether there is variation
in the mass distributions of GMCs in and out of spiral arms.
Unfortunately, the spiral arms in M33 are not nearly as well defined
as they are in the Milky Way. \citet{hs80} identified ten spiral arms
in the flocculent structure of M33 with two arms being dominant.
\citeauthor{hs80} attributed the two primary arms to spiral density
waves.  This conjecture was bolstered by observations in the near
infrared, which revealed that the spiral structure of the old stellar
population, and hence the stellar mass, was dominated by these two
``grand design'' spiral arms \citep{m33-nir}.  The remaining eight
arms are mainly located in the outer galaxy.  They were primarily
defined by the presence of young stellar associations but also
correlate well with narrow filaments of atomic gas
\citep[][EPRB]{deul}. \citeauthor{hs80} attributed these eight
secondary arms to an unknown mechanism other than density waves.

For present purposes, we have defined the locations of the M33 spiral
arms using a 3.6~$\mu$m image of the galaxy taken using the IRAC
instrument of the {\it Spitzer Space Telescope} \citep{m33-irac}.  A
small correction was made for dust emission in the 3.6~\micron\ image
leaving emission primarily from the old stellar population.  An
azimuthally-symmetric exponential disk was fit to the result.  The
spiral arms become apparent as regions of positive surface brightness
after subtracting the model disk.  Detailed analysis will be presented
elsewhere, but Figure~\ref{comap2} shows the spiral arm locations with
respect to the CO emission.

\begin{figure}
\plotone{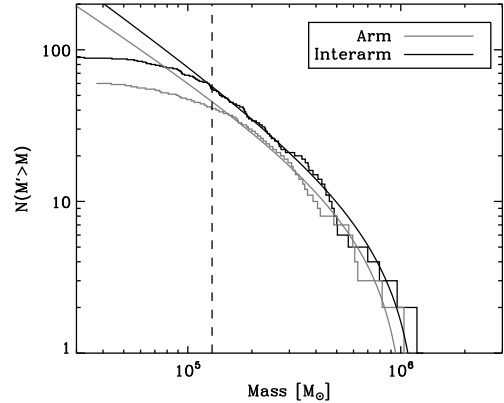}
\caption{\label{inoutarm} Comparison of the cumulative 
mass distributions for arm and interarm GMCs.  Truncated power law
fits reveal a slightly shallower underlying slope for the ``arm''
distribution, though the level of truncation in the distributions
appears similar between the two populations.}
\end{figure}

Figure \ref{inoutarm} compares the mass distributions of GMCs in and
out of the grand design spiral arms of M33.  Only slight differences
are seen.  The distributions in both arm and interarm regions show a
truncation at high mass, but massive GMCs are found in both
environments.  The truncation levels and masses are roughly equal.
Although a two-sided KS test suggests that the distributions have
different shapes, the difference is barely significant
($P_{KS}=0.04$). Unlike the sharp cutoff in the inner galaxy mass
distribution, the arm and interarm mass distributions lack a specific
feature that sets them apart. Table \ref{mspectable} gives parameters
of truncated power law fits.

\begin{figure}
\plotone{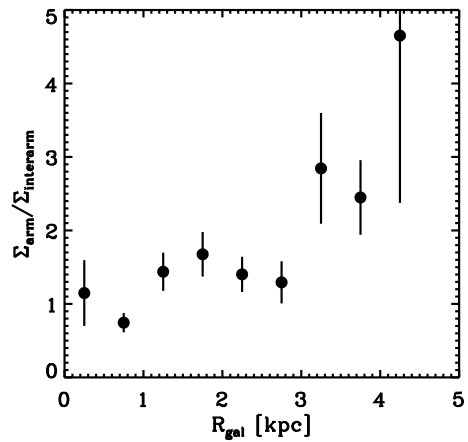}
\caption{\label{arminter} The ratio of surface densities of molecular
gas between the arm and interarm regions of M33.  The inner regions of
the galaxy show a constant value of $\sim 1.5$ but there is an apparent
increase at larger galactocentric radius.}
\end{figure}

In addition to comparing arm-interarm mass distributions of GMCs, we
examined arm-interarm variations in the column density of molecular
material.  Figure~\ref{arminter} shows the ratio of the average
surface density of CO found in spiral arms compared to that in
interarm regions as a function of galactocentric radius.  In the inner
part of the galaxy, the ratio has a roughly constant value of 1.5,
comparable to other flocculent galaxies (NGC 5055: 2--3,
\citealp{n5055-arminter}; NGC 6946: $\sim 1.4$,
\citealp{n6946-walsh}).  The value is significantly smaller than
galaxies with stronger grand-design structure such as M31, where the
ratio approaches 20 \citep{iram-m31-aa}, or M51, where contrasts of a
factor of 17 have been observed \citep{gb-m51}. In M33, much of the
``interarm'' molecular material is associated with the secondary
spiral arms identified by \citet{hs80}.  However this is a natural
consequence of star formation associated with GMCs.  Wherever GMCs
exist, one would expect newly formed stars to exist as well and create
secondary ``arms.''  The strong correlation between GMCs and high mass
stars (as traced by \ion{H}{2} regions) was established in EPRB.
Regardless of the visual appearance, in the inner regions of M33, the
CO distribution is only weakly coupled to the stellar mass
distribution, as evidenced by the small arm/interarm variation.

The increase in the arm-interarm contrast at large galactocentric
radius appears statistically significant.  However, at these
distances, the spiral density wave becomes poorly defined and may be
following enhancements in the stellar light distribution associated
with star formation in molecular clouds thereby increasing the
arm-interarm contrast.  Carefully determining the location and
amplitude of the spiral pattern present in the old stellar population
will clarify these results.

\subsection{Understanding Variations in GMC Mass Distributions}
\label{variation}
The results of \S\S \ref{gmcs-again} and \ref{massspec} demonstrate
that the mass distribution of GMCs is changing significantly across
the face of the galaxy.  The highest mass GMCs are found at
intermediate galactic radii (2~kpc~$\la R_{gal} \la$~4~kpc) and no
clouds with masses larger than $8\times 10^5~M_{\odot}$ occur outside
this region.  These changes are circumstantially related to variations
in the galaxy roughly demarcated at these radii ($\sim$2 and $\sim$4
kpc).  To generally describe the galaxy, we divide it into three
sections at small ($R_{gal}<2$~kpc), intermediate ($2<R_{gal}<4$ kpc)
and large ($R_{gal}>4$ kpc) galactic radii.  At small galactocentric
radii, the spiral structure is not as well-defined as it is for
intermediate radii \citep{hs80,m33-nir}.  At large radii, the disk of
the galaxy develops a warp \citep{cs97,cs00} and ultimately becomes
gravitationally stable \citep[$R_{gal}>6$~kpc;][]{c03}.  These
divisions between these gross features of the galaxy correspond
roughly to the regions where we see changes in the molecular cloud
mass distribution.

While this correlation is suggestive, the likely regulator of the
molecular cloud masses is structure of the atomic gas from which the
GMCs must have formed \citep{rpeb03,psp5}.  The GMCs are found at the peaks
of the atomic gas distribution (EPRB), a generic feature of the ISM in
galaxies dominated by atomic gas \citep{psp5}. In M33, the atomic gas
is distributed in what appears to be a filamentary network
\citep{deul}.  The character of this network varies with radius in the
galaxy consisting of small, relatively faint filaments in the inner
galaxy whereas there are large (kiloparsec scale) filaments in at
intermediate radii associated with the spiral arms and then large,
less bright filaments are large radius.  We quantify this description
by decomposing the bright filamentary network into a set of clouds and
examining the ``cloud'' properties.  While we adopt a similar
procedure to the segmentation that identifies GMCs, such ``clouds'' do
not necessarily represent distinct physical objects.  The segmentation
is a way to measure the characteristic sizes and masses of objects in
the atomic ISM of M33 across the galaxy.

To generate a catalog of \ion{H}{1} objects, we apply a three-tiered
brightness cut at $T_A=64,48,$ and $32$ K to the map of \citet{deul}.
Connected regions of pixels are identified at each brightness cut.
New regions at a given brightness cut are identified as new clouds.
Regions containing pixels from higher levels have those pixels
assigned to the closest predefined region.  Here ``closest'' means the
region that was identified at a higher level to which a pixel is
connected by the shortest path contained entirely within the new
region.  The algorithm is similar to CLUMPFIND \citep{clumpfind}, but
the coarse contouring levels prevent every local maximum from being
identified as as a cloud.  The shortest-path criterion is necessary
for decomposing the \ion{H}{1} emission since the ``clouds'' blend
into a continuous network at the low brightness thresholds.  Figure
\ref{deepmap} suggest that a similar criterion would be necessary for
identifying GMCs in a CO map with higher sensitivity.  We reiterate
that the decomposition is only a means to characterize the sizes of
structures in the atomic gas across the galaxy in a uniform manner.

\begin{figure}
\plotone{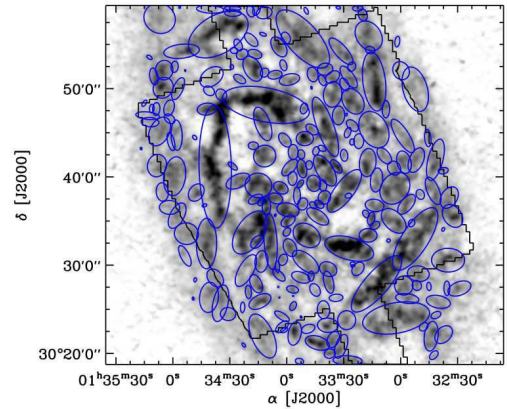}
\caption{\label{himap} Map of the integrated intensity of \ion{H}{1}
21-cm emission from \citet{deul} with locations of \ion{H}{1} clouds
indicated.  The clouds are identified based on the algorithm described
in \S \ref{variation}.  The clouds are simply convenient descriptors
of the filamentary structure of the atomic gas and do not necessarily
correspond to physically relevant objects.}
\end{figure}

In Figure \ref{himap}, we show the integrated intensity map of
\citet{deul}, restricted to the GMC survey area, with the ellipses
indicating measurements of the major ($\ell_{maj}$) and minor
($\ell_{min}$) axes of the clouds in blue.  For each \ion{H}{1} cloud
of mass $M$, we calculate the characteristic mass ($M_{char}$) of
objects that would form by gravitational instability along the
filament using the prescription of \citet{eg94}.  A major result of
Elmegreen's analysis including both spiral potential and magnetic
fields is that the fragmentation of neutral gas into ``superclouds''
(which would be substructures in the ``clouds'' we catalog) can be
evaluated in terms of the local quantities that characterize the gas,
rather than using disk-averaged values.  The characteristic mass of
these superclouds is thought to establish the upper cutoff observed in
the molecular cloud mass distribution in the Milky Way
\citep{eg94,kos03}.  In this picture, disk instabilities form atomic
superclouds which are (marginally) self-gravitating and GMC complexes
form by turbulent fragmentation and collapse inside the superclouds.
The total mass of the parent supercloud would then establish the
maximum mass of the distribution.

\begin{figure}
\plotone{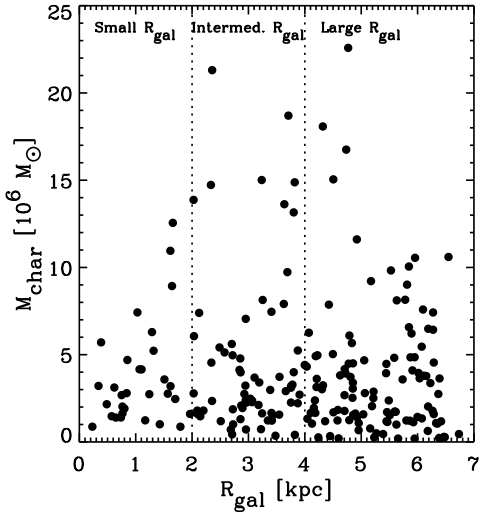}
\caption{\label{charmass} Characteristic mass of fragmentation as a
function of $R_{gal}$ for the \ion{H}{1} clouds identified in Figure
\ref{himap}.  The mass is calculated using the formula from
\citet{eg94} and the measured \ion{H}{1} cloud properties. The
characteristic mass of fragmentation peaks at intermediate $R_{gal}$
which may explain the abundance of high mass clouds in that region.}
\end{figure}

We apply the local fragmentation formalism to all the \ion{H}{1}
clouds emphasizing the principal result of the Elmegreen analysis: the
use of local conditions to evaluate the characteristic fragmentation
masses.  This application will illustrate how variations in the
cataloged atomic cloud properties would translate into variations in
the characteristic mass of superclouds and thus into the cutoffs in
the GMC mass distribution.
In terms of the
properties of the identified atomic gas clouds, the characteristic
mass of superclouds is given by \citet{eg94} as:
\begin{equation}
M_{char}=\pi \ell_{min} c_{g}\sqrt{\frac{\mu}{2G}}
\end{equation}
where $\mu$ is the mass per unit length of the clouds ($M/\ell_{maj}$)
and $c_g$ is the three-dimensional velocity dispersion of the gas,
taken to be 18 km~s$^{-1}$ for M33 \citep{cs97} assuming isotropy.
We plot the characteristic mass of supercloud formation as a function
of galactocentric radius in Figure \ref{charmass}.  The figure shows
that the atomic structures that could support fragmentation into large
mass superclouds are most abundant at intermediate values of $R_{gal}$.
Indeed, the factor-of-three variation in the maximum mass is
consistent with the idea that the mass of fragmentation controls the
variations in the maximum mass of molecular clouds seen between inner
and intermediate radii.  However, the results do not adequately
explain the abrupt fall-off in the number of GMCs seen at large
$R_{gal}$ since there are apparently massive clouds at this radius
which fulfill the instability criteria.  Other additional factors may
contribute to the dearth of GMCs at this radius.  The surface density
of the \ion{H}{1} disk becomes comparable to the stellar disk at
$R_{gal}\approx 4.5$~kpc which will result in a significant increase in the
\ion{H}{1} scale height \citep{pressure2}, not accounted for in our
simplification of \citet{eg94}.  The disk also develops a warp
\citep{cs97,cs00} at this radius though the disk remains
gravitationally unstable to structure formation out to larger
$R_{gal}$ \citep{c03}.

The locations and mass distributions of GMCs are consistent with being
governed by the structure in the atomic gas.  For $R_{gal} \la 2$ kpc
it appears that massive stars establish the structure of the ISM based
on the preponderance of wind-driven shells in the inner galaxy
\citep{ddh90}.  The GMCs in the inner galaxy appear on the edges of
these holes (EPRB) suggesting that the clouds are formed in the shells
swept up in winds from high mass stars.  At intermediate and large
(i.e.~$R_{gal}\ga 4$~kpc) galactic radii, there are substantially
fewer wind-driven shells \citep[EPRB]{ddh90}, and the molecular ISM
becomes more concentrated along the spiral arms (Figure
\ref{arminter}).  At large radii, GMCs disappear entirely, as does
significant spiral structure, and the rotation curve flattens
appreciably \citep{c03}.  The molecular gas that is present at large
radii may be the result of small scale compressions of the neutral ISM
producing less massive molecular clouds.  Given that applying simple
instability analysis to real galactic disks only leads to suggestive
correlations, resolving the variation of cloud mass distributions will
likely require sophisticated simulations of molecular cloud formation
across the galactic disk with sufficient detail to match to these
observations.  Further progress on the effects of spiral structure and
molecular cloud formation will be possible in conjunction the high
resolution map VLA+GBT map of \ion{H}{1} emission by Thilker \& Braun
(forthcoming).  Progress on simulations and observations will
facilitate the critical evaluation of the above conjectures and better
establish the association between the molecular and atomic phases of
the ISM.  The variations in the mass distributions of molecular clouds
reported in this paper will represent an important observational
feature to reproduce in theoretical analyses.

\section{Summary and Conclusions}
\label{summ}

New maps of CO ($J=1\to 0$) emission in M33, produced by combining
existing interferometric and single-dish maps, including new
single-dish observations from the NRO 45-m, give an improved catalog
of giant molecular clouds (GMCs) and a more sensitive measurement of
the non-GMC CO emission.  The GMC properties in the new catalog are
unaffected by the spatial filtering and non-linear flux recovery of
the existing interferometer catalog.  The relative calibrations of all
three data sets agree, and the total flux in the merged and combined
maps match the FCRAO map in the regions of overlap.  The GMC
properties in the new catalog are corrected for instrumental effects
using the method of \citet{props} resulting in superior estimates of
the cloud properties.

Improvements in the estimates of cloud properties as well as a high
sensitivity map of the inner galaxy yield several new results.  The
fraction of molecular gas found in GMCs ($M>1.3\times10^5~M_{\odot}$)
declines with galactocentric radius, ranging from 60\% at $R_{gal}=0$
to 20\% at $R_{gal}=4.0$~kpc.  Based on the higher sensitivity
BIMA+FCRAO+NRO map, molecular gas that is not associated with GMCs is
located in diffuse and filamentary structures around the GMCs.  Nearly
90\% of the diffuse emission is found within 100 pc projected
separation of a catalog GMC boundary.  Beyond 4.0~kpc, the GMC mass fraction
cuts off sharply, though molecular gas is detectable to the edge of
the surveyed region at $R_{gal}=5.5$~kpc.  Even within the area
$R_{gal}<4$~kpc, where GMCs are found, the mass distribution in the
inner galaxy ($R_{gal}<2.1$~kpc) is significantly different from the
mass distribution in the outer galaxy.  In the annulus $2.1 < R_{gal}
< 4.1$~kpc, the GMCs have a power-law mass distribution of
$dN/dM\propto M^{-2.1}$.  Inside 2.1~kpc, the GMC mass distribution is
truncated at high mass ($\gtrsim 4\times 10^5~M_{\odot}$), but lower
mass clouds appear to follow the same distribution in both regions.
We argue that variations in the \ion{H}{1} gas structure are
responsible for the observed changes in the GMC mass distribution.

The southern (receding) half of the galaxy shows a truncated mass
distribution when compared to the northern half of the galaxy.  There
is no obvious difference between the mass distribution of GMCs
associated with the grand design spiral arms of the galaxy and those
GMCs in the interarm region.  In molecular surface brightness, M33
shows a arm-interarm contrast of 1.5, typical of other flocculent
galaxies.

\acknowledgements 
We are grateful for the opportunity to acquire data on the 45-m
telescope at the Nobeyama Radio Observatory (NRO). NRO is a division
of the National Astronomical Observatory of Japan under the National
Institutes of Natural Sciences.  We thank Mark Heyer for providing the
FCRAO data on M33 used in this work.  ER's work is supported by a
National Science Foundation Astronomy and Astrophysics Postdoctoral
Fellowship (NSF AST-0502605).


\facility{NRO/BEARS, FCRAO, BIMA}




\end{document}